\documentclass[aps,prl,reprint,
amsmath,amssymb]{revtex4-1}
\usepackage{graphicx}% Include figure files
\usepackage{dcolumn}% Align table columns on decimal point
\usepackage{bm}% bold math
\usepackage{amssymb}
\usepackage{tabularx}
\begin{document}

\title [Short title]%{Demonstration of Quantum CNOT-A Gate using magneto-switch}
{ Study of Microwave Assisted CNOT Gate}
\author{M. K. Mehta} % \and Pratima Sen \and J.~T.~Andrews*\and}
%\maketitle
\affiliation{Laser Bhawan, School of Physics,\\ Devi Ahilya University, Khandwa Road, Indore - 452 007 India.}
%\affiliation{Applied Photonics Laboratory, Department of Applied Physics, \\ 
%Shri G S Institute of Technology \& Science, Indore - 452 003 India.}
\author{P. Sen}%
\email[Corresponding author email: ]{pratimasen@gmail.com}
\affiliation{Laser Bhawan, School of Physics,\\ Devi Ahilya University, Khandwa Road, Indore - 452 007 India.}

\author{J. T. Andrews}
\email[]{jtandrews@sgsits.ac.in}
\affiliation{Applied Photonics Laboratory, Department of Applied Physics, \\ 
Shri G S Institute of Technology \& Science, Indore - 452 003 India.}

%\date{Received: date / Accepted:date}
\begin{abstract} 
Population evolution in a magnetic impurity doped semiconductor quantum dot has been studied by applying a sequence of pulses of chosen pulse area. By optical
excitation mechanism, the population in \(J_z=+3/2\), 
heavy hole state of valence band is carried over to 
\(J_z=-3/2\), valance band state, via the \(J=+1/2\)
conduction band states. The injected
microwaves entangle conduction band states. This arrangement is successfully employed to ascertain quantum
CNOT  operation, and the calculation predicts
maximum fidelity of 80\% for the CNOT operation. 
\end{abstract}
\maketitle
\section{Introduction}

The reduced size and dimensionality of Quantum Dots (QDs) cause confinement of electrons, therein lead to marvel phenomena like sharpening of density of states, atom like discrete electronic energy level structure, etc. The leeway of the electrons in QDs have opened up the possibilities of novel applications in quantum computation \cite{loss1998quantum, divincenzo2000physical, puri2017universal, mills2019shuttling, wenz2019quantum} which satisfy most of the fundamental requirements for realization of quantum computing viz., (i)~well characterized qubits (ii)~initialization of the states (iii)~long decoherence time (iv)~ability of forming quantum gates and (v)~qubit specific measurement capability.

Of late, coherent control and CNOT gate operation in semiconductor QDs exhibiting excitonic and biexcitonic features were reported by the authors' group \cite{qureshi2008all}. The Gate operation in those situations were restricted by the dephasing time $T_2$ (sub ps) of the quantum dot. This limitation can be overcome by using spin states as the qubits due to longer spin relaxation time. For the case of semiconductor quantum dots, the spin flipping requires large magnetic fields while in diluted magnetic semiconductor (DMS) quantum dots, moderately low magnetic fields can lift the degeneracy of spin states. Also, the above mentioned criteria of quantum computing are fulfilled by the quantum two- level system representing polarization sensitive spin states in valence and conduction band \cite{awschalom2013semiconductor}. The splitting of the valence and conduction band states are qualitatively identical to normal spin splitting in semiconductors. In case of DMS, the splitting energy is large enough to isolate the effect of environment on the transitions amongst the spin split states. The four-fold degeneracy of the heavy (\textit{hh}) and light (\textit{lh}) hole valance band states ($ J=\frac{3}{2}, {J}_{z}=\pm{\frac{3}{2}}$ and ${J}_{z}=\pm{\frac{1}{2}}$) can be lifted by applying magnetic field normal to the growth axis to ensure that the \textit{hh} splitting is larger than \textit{lh} splitting. Hanson et al, reported initialization and single shot read-out of the spin state in semiconductor quantum dots \cite{hanson2004electron}, where they fabricated a double dot device with integrated electrometer served as a two-qubit circuit. Fujita et al \cite{fujita2019angular}, measured the transfer of angular momentum, from circularly polarized photon to an electron spin state in quantum dot. Koppens et al \cite{koppens2006driven} demonstrated the feasibility of operating single-electron spin in a quantum dot as a quantum bit. Press et al \cite{press2010ultrafast} suggested that the spin of a single electron confined in a semiconductor quantum dot forms a promising qubit that may be interfaced with a photonic network. Same group demonstrated the optical initialization, rotation by arbitrary angle and projective measurement of an electron spin in a quantum dot \cite{press2008complete}. The experiments carried out by Gupta et al \cite{gupta2001ultrafast} show that optical tipping pulses can enact substantial rotations of electron spins through a mechanism dependent on the optical Stark effect in semiconductor quantum wells. In their experiment, use of prototype sequence of two tipping pulses showed reversible rotations which established the coherent nature of the tipping process. One can therefore expect that the $\pi$-pulses can be constructed to coherently control spins in semiconductors on femto-second time scales. Wei et al \cite{wei2014universal} proposed Universal quantum gates on electron-spin qubits with quantum dots inside single-side optical microcavities. Rosenblum et al \cite{rosenblum2018cnot} proposed controlled-NOT (CNOT) gate between two multiphoton qubits in two microwave cavities. Castelano et al \cite{castelano2018optimal}, proposed the set of universal quantum gates, based on the quantum optimal control theory driven by applied electric fields to semiconductor double quantum dots in semiconductor nanowire.

A deterministic and scalable scheme to construct a two-qubit CNOT gate and realize entanglement swapping between photonic qubits using a QD spin in a
double-sided optical microcavity was demonstrated by Wang et al \cite{wang2013deterministic}. Plantenberg et al \cite{plantenberg2007demonstration} also demonstrated CNOT operations
on a pair of superconducting quantum bits. They used microwave pulses of appropriate energy to a
single pair of coupled qubits. Imamoglu et al \cite{imamog1999quantum} proposed a scheme in which controlled interactions between two distant quantum dot spins is mediated the vacuum field of a high finesse microcavity. They have shown that the Raman transitions induced by classical laser fields and the cavity-mode can be used to make controlled-not operations and arbitrary single qubit rotations can be realized.

In the present communication, we propose a new scheme for quantum information processing based on electron spin in a DMS-QD coupled through a microwave field.
 We have used timed pulse sequences to control electron spin dynamics. The first pulse is a left circularly polarized radiation from a laser, which excites electron from ${J}=|\frac{3}{2}\rangle$ state in valence band to generate population in the ${J}=|\frac{1}{2}\rangle$ state in conduction band. The second pulse of energy equals to the difference in the energies of spin split states from a microwave  source is applied within a time much shorter than the dephasing time of the spin state and leads to the occupation probability of electron in ${J}=|-\frac{1}{2}\rangle$ state in conduction band. The spin flip electron is then de-excited by applying a third pulse from the same laser source but is a right circularly polarized pulse of energy equals to the transition energy between ${J}=|-\frac{3}{2}\rangle$ state to ${J}=|-\frac{1}{2}\rangle$ state.

\section{Theoretical formulations}

\begin{figure}
    \centering
\includegraphics[width=.85\columnwidth]{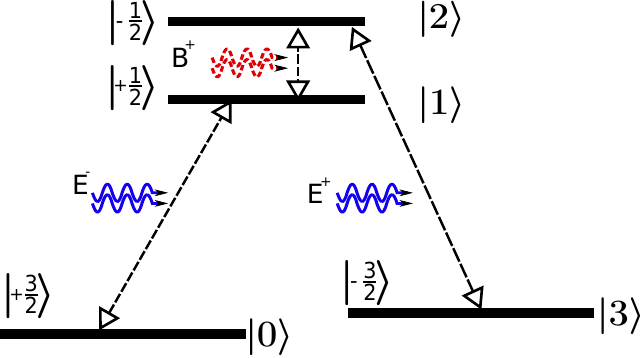}
\caption{Schematic diagram of the transition processes of two coupled 2-level systems.}\label{fig1}
\end{figure}

Atom like discrete electronic spectra provides the base of qubits in quantum dots. In semiconductors QDs, the transition between $|\pm 3/2 \rangle$ valence band states and $|\pm 1/2 \rangle$ conduction band states are sensitive to the nature of optical polarization of the exciting laser. The degeneracy of these states can be lifted by the application of suitable magnetic field. In magnetic impurity doped semiconductor quantum dots,  it is easy to achieve spin splitting of these states at moderately large applied magnetic field. As shown in Figure \ref{fig1}, the spin splitting leads to the creation of two pairs of dipole allowed transition energy levels $|+3/2\rangle \rightleftharpoons |+1/2\rangle$ and $|-3/2 \rangle\rightleftharpoons|-1/2\rangle$. The pair of these levels can be sought as two qubits.

The present article, aims at designing controlled-NOT operation \cite{monroe1995demonstration, o2003demonstration}  by coupling these states via strong laser fields and entangling by $(|+1/2\rangle \rightleftharpoons |-1/2\rangle)$ in the conduction band by application of microwave field similar to electron spin resonance (ESR) technique. 

\begin{figure}
    \centering
 \includegraphics[width=.85\columnwidth]{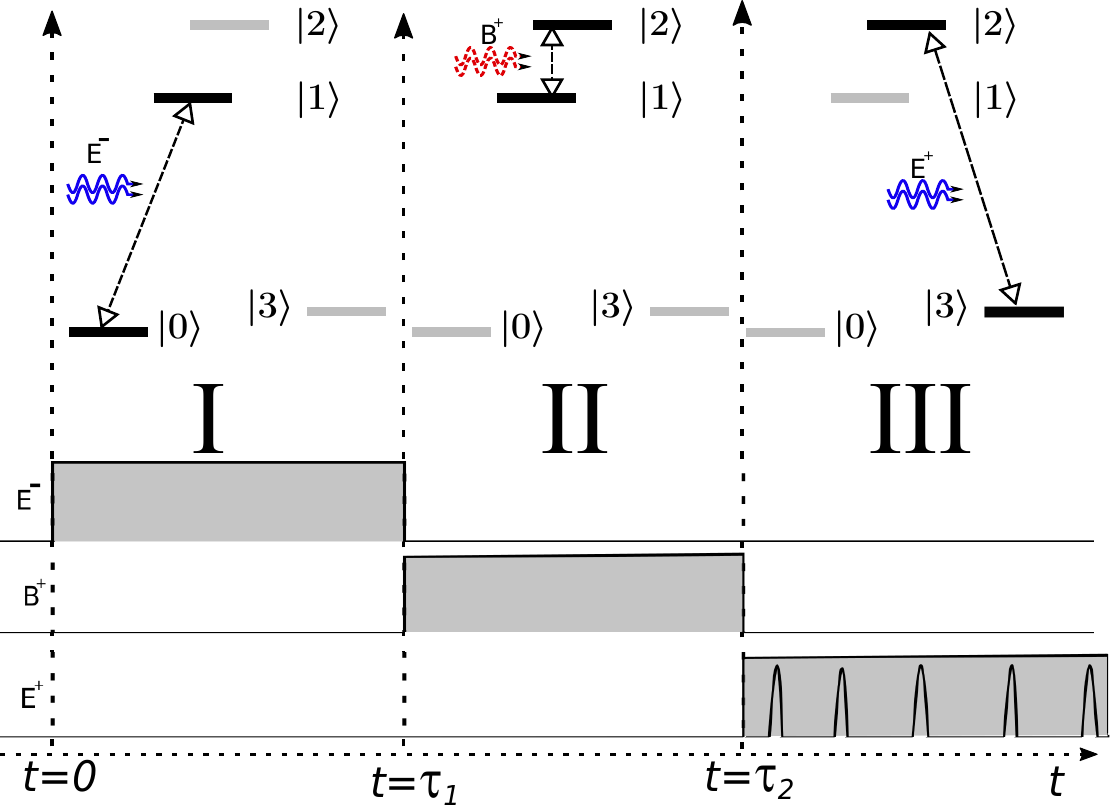}
\caption{Schematic representation of energy level
diagram in each regime's electromagnetic pulses vs time. Regime-I, is governed through left circularly polarized $E^-$ excitation pulse, which raises the electron from $|0\rangle$ to $|1\rangle$. Regime-II, is governed through the microwave  $B^+$  pulse which raise the electron from $|1\rangle$ to $|2\rangle$. Regime-III, is governed further by right circularly polarized $E^+$ pulse which depicts the transition of the electron from $|2\rangle$ to $|3\rangle$. The spikes in regime-III, show the repetition of pulsed laser for the CNOT operation.   
The pulse switch on-off times are shown at the
lower part of the diagram. }\label{regimes}
\end{figure}

In Figure \ref{fig1}, we represent the  valance band (VB) ($|+3/2\rangle$ and $|-3/2\rangle$) states
as $\left|0\right\rangle$ and $\left|3\right\rangle $, whereas, the conduction band (CB) $|+1/2\rangle$ and $|-1/2\rangle$ states
as the $\left|1\right\rangle$ and $\left |2\right\rangle$, respectively
so as to simplify the representations. We follow a time sequence of three pulses. At the onset, the first pulse which is left circularly polarized pump pulse, 
raises the population from $|0\rangle $  to $|1\rangle$.
This we call as regime-I. The second pulse in regime-II, is a microwave field with an energy matching with the spin split energy of CB states, induces single-spin rotation in the quantum dot \cite{loss1998quantum, burkard1999coupled, elzerman2004single, koppens2006driven, press2010ultrafast} and raises the population from $|1\rangle$  to $|2\rangle$. Subsequently, in regime -III, the third pulse which is right circularly polarized electromagnetic pulse, de-excites population from $|2\rangle$ to $|3\rangle$ assuming that state $|3\rangle$ remains unoccupied in magnetic impurity doped SQDs. The schematics of these processes are shown in Fig.~\ref{regimes}. During the excitation by the optical pulse of duration $\tau_1$ in regime - I $(0\leq t \leq \tau_1)$, coherent oscillations of population occur within $\left|0\right\rangle$ and $\left|1\right\rangle$ states. For the regime -II $(\tau_1 \leq t \leq \tau_2)$, the microwave pulse of duration $\tau_2$ is switched on such that coherent oscillations take place between $|1\rangle$ and $|2\rangle$ states
similar to that
of an ESR experiment. For times $t > \tau_2$, the optical third pulse resonantly deexcites population from state $|2\rangle$ to $|3\rangle$. Within the coherence regime, oscillations of population take place between $|2\rangle$ and $|3\rangle$.  One needs to monitor the sequencing of the pulses in such a manner that a CNOT gate operation can be realized. In Fig. \ref{regimes} , we have shown the schematic of the excitation process in time scale along with the square pulses displayed in gray color at the bottom of the figure. In regime-III, the excitation pulse may be chosen as train
of square or  Gaussian pulses as displayed in the figure. For two-qubit representation, which is a prerequisite for CNOT operation, we shall consider the present decimal numbered state representation to binary number states as $|0\rangle \Rightarrow |00\rangle$, $|1\rangle \Rightarrow |01\rangle$, $|2\rangle \Rightarrow |10\rangle$ and  $|3\rangle \Rightarrow |11\rangle$.  
In the first and third regimes, under effective mass approximation for the dipole allowed optical transitions $|0 \rangle$ $\rightleftharpoons$ $|1 \rangle$ and $|2 \rangle$ $\rightleftharpoons$ $|3 \rangle$, the corresponding interaction Hamiltonian $H'$ is given by
 
\begin{equation}
H'= -\frac{1}{2}({\mu }^{\pm}E^{\mp}+ c.c.);  \label{Ham}
\end{equation} 
where ${\mu }^{\pm }={\mu }_x\pm i{\mu }_y$ and $E^{\pm }=E_x\pm iE_y$. We use density matrix approach and in particular, Bloch function technique to examine the time evolution of population in different states for the complete time sequencing described above. 

The density operator $\rho (t)$ corresponding
to the radiation-matter interactions under regime
- I, II and III are determined for the transition between two-level system using the Liouville-von Neuman equation  \cite{kloeffel2013prospects, steck2007quantum},
\begin{equation}
\dot{\rho }(t)=-\frac{i}{\hbar }\left[H\left(t\right),\rho \left(t\right)\right]-\Gamma
\rho(t),\label{rho}
\end{equation}
where $\Gamma$ is the phenomenological decay parameter.
The  Hamiltonian $H(t) (=H_0 +H'(t))$ and density matrix $\rho(t)$ are expressed as
\begin{subequations}
\begin{equation}
\rho(t)=\begin{pmatrix}
\rho_{00} & \rho_{01} &\rho_{02}&\rho_{03} \\ 
\rho_{10} & \rho_{11} &\rho_{12}&\rho_{13} \\ 
\rho_{20} & \rho_{21} &\rho_{22}&\rho_{23} \\ 
\rho_{30} & \rho_{31} &\rho_{32}&\rho_{33}
\end{pmatrix},
\end{equation}
\begin{equation}
H_0=\hbar\begin{pmatrix}
\omega_{00}&0&0&0\\
0 & \omega_{11} & 0 & 0\\
0&0&\omega_{22}&0\\
0&0&0&\omega_{33}
\end{pmatrix},
\end{equation}
and
\begin{equation}
H'(t)=\begin{pmatrix}
0&\Omega^{+}_{01}(t)&0&0\\
\Omega^{+}_{10}(t) &0 & \Omega_{12}(t) & 0\\
0&\Omega_{21}(t)&0&\Omega^{-}_{23}(t)\\
0&0&\Omega^{-}_{32}(t)&0
\end{pmatrix}.
\end{equation}\label{rh}
\end{subequations}
Here, we defined $\rho_{ij} = |i\rangle \langle j|$ and $\hbar\Omega_{ij}= \langle i\left|H^{'}\right|j\rangle$. The radiation-matter interaction are defined via the Rabi flopping frequencies as:\\
\begin{subequations}
\begin{eqnarray}
\hbox{Regime I:} \hspace{10mm}&\Omega^{+}_{01}(t)&=\frac{\mu_{01}^{+}E^{-}(t)}{\hbar},\\
\hbox{Regime II:} \hspace{10mm}&\Omega^{}_{12}(t)&=\frac{\wp_{12}B(t)}{\hbar},\\
\hbox{Regime III:} \hspace{10mm}&\Omega^{-}_{23}(t)&=\frac{\mu_{23}^{-}E^{+}(t)}{\hbar},
\label{Omega}
\end{eqnarray}
\end{subequations}
with $\mu_{ij}$ and
$\wp_{ij}$ are the electric and magnetic transition dipole moments, respectively and the
Rabi flopping frequencies satisfy the relation,
$\Omega^{\pm}_{ij}= (\Omega^{\pm}_{ji})^*$.

According to the time sequencing of the pulses, one can find
\begin{equation}
\Omega_{ij}(t)=
\begin{cases}
\Omega_{01}^{+}\neq0,\;\Omega_{12}=0,\; \Omega_{23}^{-}=0  & \text{for}\;\; 0\leq t\leq \tau_{1},\\
\Omega_{01}^{+}=0,\;\Omega_{12}\neq0,\; \Omega_{23}^{-}=0 & \text{for}\;\; \tau_1\leq t\leq \tau_{2},\\
\Omega_{01}^{+}=0,\;\Omega_{12}=0,\; \Omega_{23}^{-}\neq0  & \text{for}\;\; \tau_2\leq t\leq T_{1}.
\end{cases}\label{Ome}
\end{equation}

We arrange the proposed density matrix in terms of the Bloch vector 
$u(t)={\rho}_{ij}(t)+{\rho}_{ji}(t)$,
$v(t)=i({\rho}_{ji}(t)-{\rho}_{ij}(t))$ and
$w(t)={\rho}_{ii}(t)-{\rho}_{jj}(t)$. The subscripts $i$ and $j$ are appropriately chosen as,
        $ij \rightarrow 01$ %   \rightarrow |0\rangle$ and         $j  \rightarrow |1\rangle$
       in regime-I,
        $ij \rightarrow 12$ % \rightarrow |1\rangle$ and         $j  \rightarrow |2\rangle$
 in regime-II, and 
        $ij\rightarrow 23$ %   \rightarrow |2\rangle$ and         $j  \rightarrow |3\rangle$
       in regime-III.
 
Using equations (\ref{Ham}) - (\ref{Ome}), we get the 
Bloch equations as 

\begin{subequations}
\begin{equation} 
\dot{u}+\mathrm{\Delta }v+\frac{u}{T_2}=0, 
\end{equation} 
\begin{equation} 
\dot{v}-\mathrm{\Delta }u-\Omega w+\frac{v}{T_2}=0, 
\end{equation} 
\begin{equation} 
\dot{w}+\Omega v+\frac{w-w^0}{T_1}=0.
\end{equation} 
\end{subequations}
The solutions to the Bloch equations as instructed by Torrey \cite{torrey1949transient}, are of the form:
\begin{equation}
M\left(t\right)=Ae^{-at}+Be^{-bt}{\mathrm{cos} \left(st\right)\ }+Ce^{-bt}{\mathrm{sin} \left(st\right)\ }+D,
\end{equation} 
where $M\left(t\right)$ the solution for $u\left(t\right),\ v(t)$ and $w(t)$. Here, $a$, $b$, and $s$ are functions of detuning
parameter ($\Delta)$, Rabi flopping frequency  $(\Omega)$, and relaxation  parameters
($T_1,\; T_2$). $A$, $B$, $C$, and $D$ are constants, where $D$ shows the steady state behavior. It is toilsome to obtain a generalized solution to these equations. Under the assumptions, $(i)~T=T_1=T_2,$ and $(ii)~\Omega \gg \frac{1}{T_1},\; \frac{1}{T_2}$, the solutions for Block vector components are \cite{harper1977nonlinear} 
\begin{subequations}
\begin{eqnarray}
u(t)&=& e^{-\frac{t}{T}} \left\{u(0)-{\Delta \left [v(0)-\xi\right ]{\frac{\sin({\beta}{t})}{\beta}}}\right.\nonumber\\
     && \left. + \Delta \left [{\Delta}u(0)+ {\Omega}{w(0)}-\frac{\xi}{T}\right ]\right.\nonumber\\
     && \left.\times{\frac{\left(\cos({\beta}{t})-1\right)}{\beta^2}} +\Delta\xi{T}\right\}-\Delta\xi{T},\label{u1}\\
v(t) &=& e^{-\frac{t}{T}} \left\{{ \left [v(0)-\xi\right ]{\cos({\beta}{t})}}\right.\nonumber\\
     && \left. + {\left [{\Delta}u(0)+ {\Omega}{w(0)}-\frac{\xi}{T}\right ]{\frac{\sin({\beta}{t})}{\beta}}}\right\} +\xi,\label{v1}\\
w(t) &=& e^{-\frac{t}{T}} \left\{w(0)-w^0-{\Omega \left [v(0)-\xi\right ]{\frac{\sin({\beta}{t})}{\beta}}}\right.\nonumber\\
     && \left. + {\Omega \left [{\Delta}u(0)+ {\Omega}{w(0)}-\frac{\xi}{T}\right ]{\frac{\left(\cos({\beta}{t})-1\right)}{\beta^2}}}\right.\nonumber\\
     && \left. +\Omega\xi{T}\right\}+w^0\left(1-\frac{\Omega\xi{T}}{w^0}\right).\label{w1}
\end{eqnarray}\label{uvw1}
\end{subequations}
Where, $\xi$ is dimensionless parameter defined as $\frac{{\Omega w^0}/{T}}{\Omega^2+\Delta^2+\frac{1}{T^2}}$.
Also, $T_1$, and $T_2$ are the corresponding recombination and dephasing times related to the damping parameter $\Gamma$ $\sim$ $(1/{T_1}+1/{T_2})$. $u(0)$, $v(0)$ and $w(0)$ are the values of the Bloch vector at the beginning of the specific regime.  The detuning parameter representing the difference between the energy of excitation pulse and transition energy
is redefined as $\Delta_{ij}=\omega -(\omega_{jj}-\omega_{ii})$. $\beta=\sqrt{{\Omega^2}+{\Delta^2}}$, and $w^0$ is the source term where electrons enter quantum states via other means.
 We now apply the solution given by equations (\ref{uvw1}) to the three regimes described above.
 
For regime-I, we choose the excitation pulse to be left circularly polarized such that the single electron is raised from $|0\rangle $ to $|1\rangle$ (or $|+{\frac{3}{2}}\rangle$ to $|+{\frac{1}{2}}\rangle$). The duration of pulse is chosen to be shorter than the dephasing time and the amplitude of the pulse is taken such that the Rabi frequency $\beta_{ij}$ exceeds the inverse of dephasing time. Consequently, optical nutation takes place within the pulse duration and the population oscillates between the excited and the ground state.
\begin{figure}
    \centering
\includegraphics[width=.85\columnwidth]{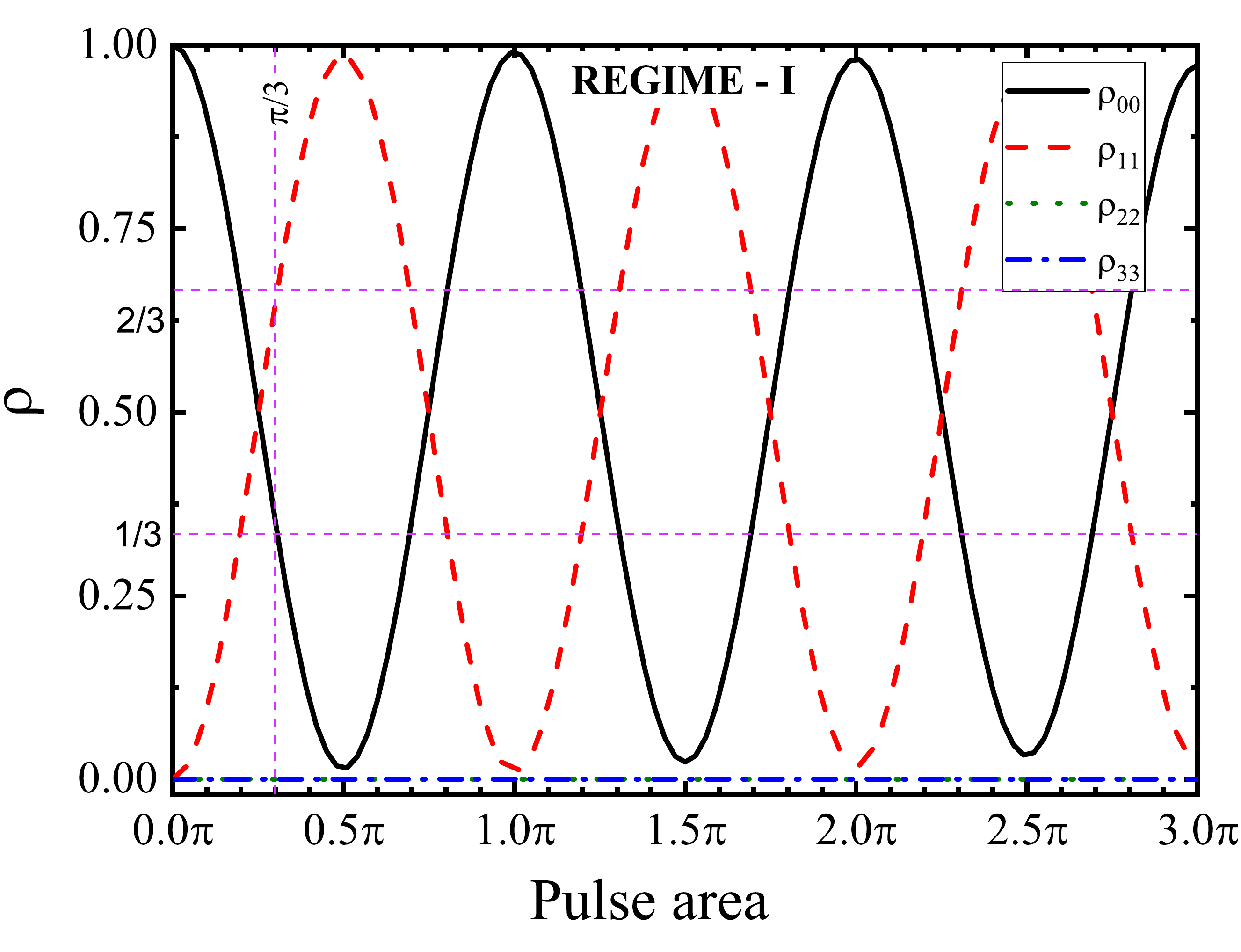}
\caption{Evolution of diagonal elements of density
matrix $\rho$ with  pulse area of %($\beta t$)
the electromagnetic radiation in regime
- I. If a $\pi/3$ pulse is applied the states
$|0\rangle$ and $|1\rangle$ are prepared with
a probability of 1/3 and 2/3, respectively.}\label{fig2}
\end{figure}

As discussed earlier, for the first regime ranging from $t=0$ to $t=\tau_1$,  %and in Fig.~\ref{regimes},
the initial conditions for the Bloch vector are
at $t=0$, $u(0)=0$, $v(0)=0$ and $w(0)=-1$. Accordingly,
 eqs. (\ref{uvw1}) reduces to

\begin{subequations}
\begin{eqnarray}
u(t)&=&\frac{\Omega_{01}^{+}\Delta_{01}}{\beta^{2}_{01}}(1-\cos(\beta_{01}
t))e^{-\frac{t}{T_2}},\\
v(t)&=&-\frac{\Omega_{01}^{+}}{\beta_{01}}\sin(\beta_{01}
t)e^{-\frac{t}{T_2}},\\
w(t)&=&\left[1+\left[\frac{\Omega_{01}^{+}}{\beta_{01}}\right]^{2}(1-\cos(\beta_{01}
t))\right]e^{-\frac{t}{T_2}}.
\end{eqnarray} \label{uvwreg1}
\end{subequations} 
 
In the forthcoming analysis, we have strictly
restricted ourselves to coherent regime,  such
that $T =T_2$.  %discussion, we have   this regime, the pulse  will not affect the population in the state $|2\rangle$ and $|3\rangle$. 
Consequently, the time evolution of the population in the states $|0\rangle$, and $|1\rangle$ are obtained from eqs. (\ref{uvwreg1}). 

 According to the area theorem which plays key role in quantum operations corresponding to transition in a qubit, the pulse area is given by $ \frac{1}{\hbar}\int_{0}^{t} \overrightarrow{\mu}.\overrightarrow{E}(t) dt$. Here, $t$ is the pulse duration, $\mu$ is transition dipole moment and
 $E(t)$ is the time dependent amplitude of the pulse under resonance condition, $\Delta$ is zero and in the first regime the pulse area is ${\beta_{1}t_{p}}/{2\pi}$.
In the coherent transient regime
the pulse area corresponds to sub-picosecond time duration. In Fig. \ref{fig2} we have plotted population density as a function of pulse area which shows the oscillations between $|0\rangle$ and $|1\rangle$. From the figure one can notice that the probability of occupation in level $|1\rangle$ can be controlled by selection of the proper pulse area.  Our objective is to monitor the population in the $|1\rangle$ state so that it can be excited to the next level $|2\rangle$ in regime-II.

A second pulse can
be applied at an appropriate time such that the
fidelity of the CNOT operation can be maximized. We choose an arbitrary time ${\tau}_{1}$ at which the first pulse is switched off and the second pulse is switched on. This regime is henceforth called as regime-II which lasts up to time ${\tau}_{2}$. The second pulse is a microwave pulse of energy equals to the difference in the energies of spin split states and is applied within a time much shorter than the dephasing time of the spin state. The initial conditions for the second regime are given
in Table -$1$.

\begin{figure}
    \centering
\includegraphics[width=.85\columnwidth]{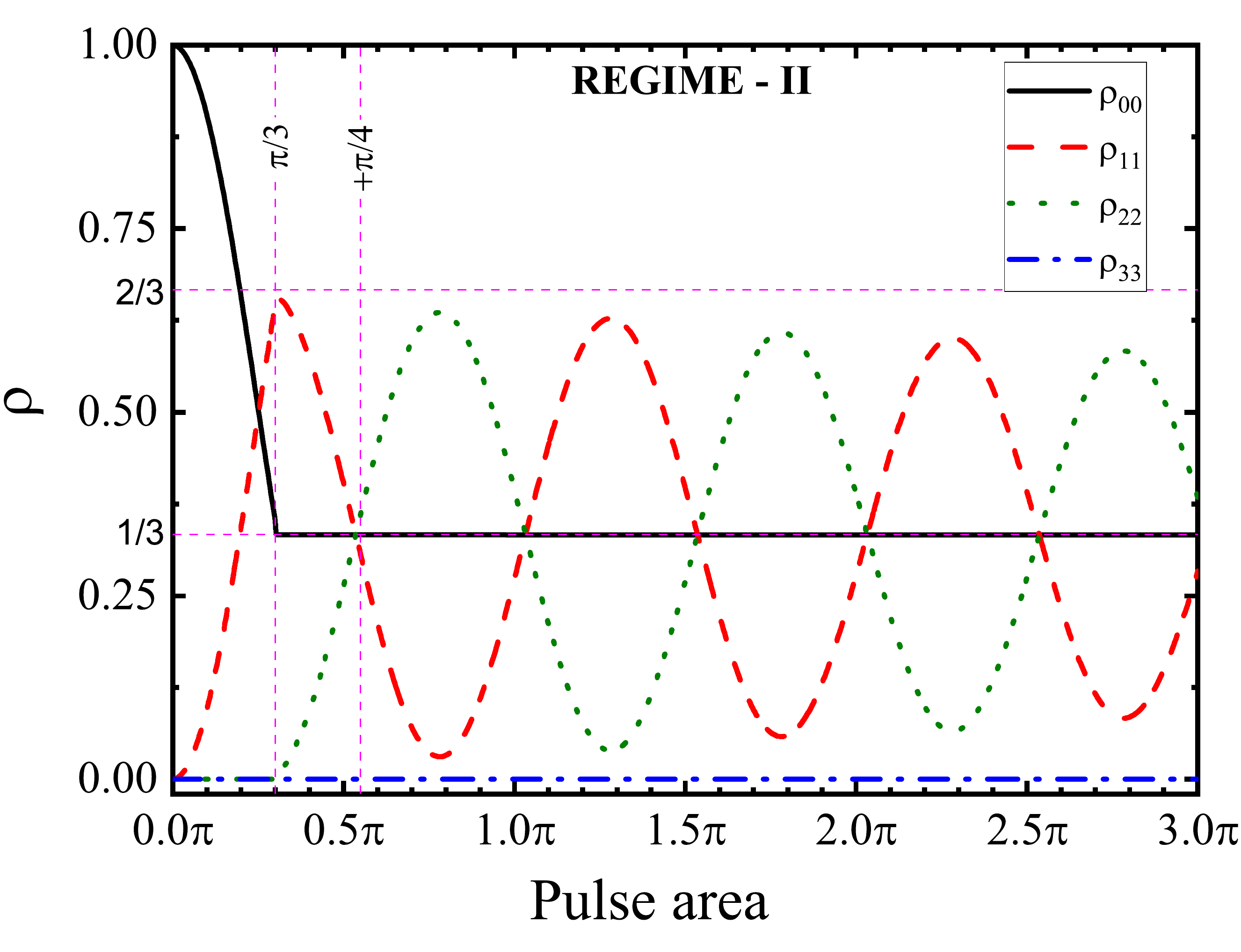}
\caption{Dependence of diagonal elements of density
matrix $\rho$ with pulse area $({\beta}t)$. If
microwave radiation is switched off after an pulse area of
$\pi/4$, the states $|1\rangle$ and $|2\rangle$, shall
be prepared with a probability of $1/3$ in each states.
   }\label{fig3}
\end{figure}

It is worth mentioning that the total population is should
be normalized to $1$ such that at the onset of the second pulse, the population in state $|1\rangle$ is given by $({1+w(\tau_{1}))}/{2}$. Consequently, for second regime,
$w'(0)^{}=-({1+w(\tau_{1}))}/{2}$. The time evolution of the Bloch vector is given by,
\begin{subequations}
\begin{eqnarray}
u(t) &=& \frac{w'(0)}{2}(\cos(\beta_{21}t)-1)e^{-\frac{t}{T_2'}},\\
v(t) &=& -w'(0)\sin(\beta_{21}t)e^{-\frac{t}{T_2'}}  \\
w(t) &=&w'(0)\left[\frac{1-\cos(\beta_{21}t)}{2}-\sin(\beta_{21}t)\right]e^{-\frac{t}{T_2'}}. \label{w2}
\end{eqnarray}
\end{subequations}
In obtaining above equations we have assumed that in presence of the pulse, the transition frequency is Zeeman shifted such that $\Delta$ in equation (\ref{uvw1}) is replaced by $\Delta\rightarrow \Delta +\Omega_{21}$ and have neglected the population from other states $(\xi=0)$. Also, $T_2'$ represents the dephasing time for $|1\rangle$ $\rightleftharpoons$ $|2\rangle$ transitions. In this regime, the population in $|0\rangle$ state is not affected and will remain nearly
constant within the coherence time. The initial population available in state $|1\rangle$ will further execute
oscillations between states $|2\rangle$ and $|3\rangle$. As an example, we have chosen the pulse area in the first regime to be  $\pi/3$ and the corresponding oscillations of populations are displayed in Fig. \ref{fig3}.  Our objective is to carry over the oscillations between $|2\rangle$ and $|3\rangle$ states via
a right circularly polarized  pulse in resonance with $|2\rangle \rightleftharpoons |3\rangle$ transition so as to follow the spin conservation. The amplitude of oscillation will depend upon the availability of population in state $|2\rangle$ at the time of excitation by the third pulse in regime-III.
The initial condition in the third regime is $w''(0)=-\frac{1+w(\tau_{1})}{4}$ (table $1$). This initial condition make
sure that the population is equally distributed in states $|1\rangle$ and $|2\rangle$. 

\begin{table}
\caption{Time evolution of Bloch vectors and the
initial conditions in each regime.}
\scalebox{0.6}{
\begin{tabular}{|c|c|c|c|c|c|c|c|c|}\hline
(1) &(2)& (3) &(4)&(5)&(6)&(7) &(9)&(10)\\\hline
regimes&time  & $u(t)$ & $v(t)$ & $w(t)$& $u(0)$ & $v(0)$ & $w(0)$ &Rabi
Freq.\\\hline
I&$0\leq t\leq \tau_1$ & ${\rho}_{01}+{\rho}_{10}$ & $i({\rho}_{01}-{\rho}_{10})$ & ${\rho}_{11}-{\rho}_{00}$ &0&0&$-
1$&$\Omega_{01}^+$\\\hline
II&$\tau_1\leq t\leq \tau_2$ & ${\rho}_{12}+{\rho}_{21}$ & $i({\rho}_{12}-{\rho}_{21})$ & ${\rho}_{22}-{\rho}_{11}$&$0$&$0$&${\rho}_{22}(\tau_1)-{\rho}_{11}(\tau_1)$&$\Omega_{12}$ \\\hline
III&$\tau_2\leq t<T_{2}$ & ${\rho}_{23}+{\rho}_{32}$ & $i({\rho}_{23}-{\rho}_{32})$ & ${\rho}_{33}-{\rho}_{22}$& $0$&$0$&${\rho}_{33}(\tau_2)-{\rho}_{22}(\tau_2)$ &$\Omega_{23}^-$\\\hline
%{\rho}_{01}(\tau_1)+{\rho}_{10}(\tau_1)
%i({\rho}_{01}(\tau_1)-{\rho}_{10}(\tau_1))
%{\rho}_{12}(\tau_2)+{\rho}_{21}(\tau_2)
%i({\rho}_{12}(\tau_2)-{\rho}_{21}(\tau_2))
\end{tabular}
}\label{regT}
\end{table}

\begin{figure}
    \centering
\includegraphics[width=.8\columnwidth]{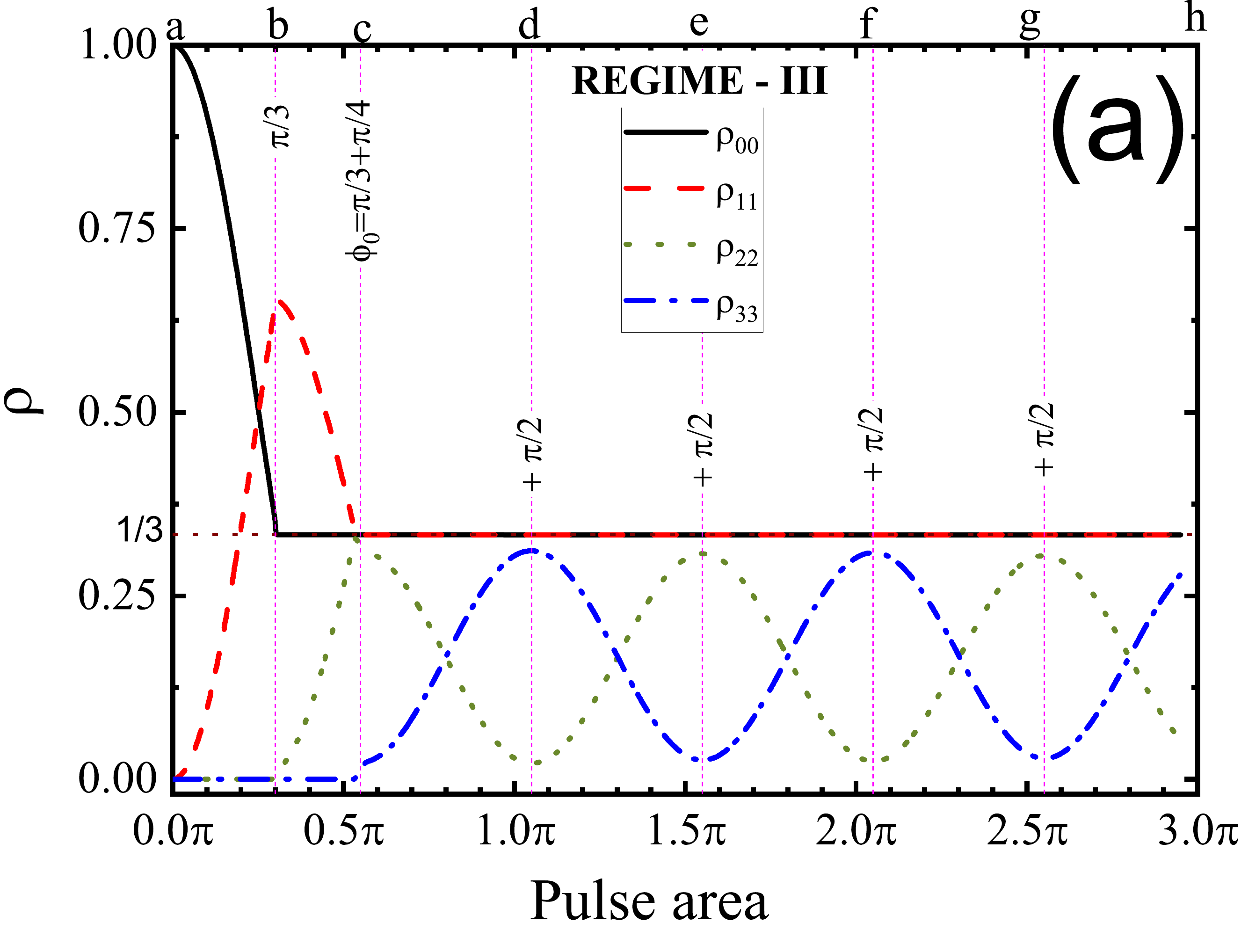}
\includegraphics[width=.8\columnwidth]{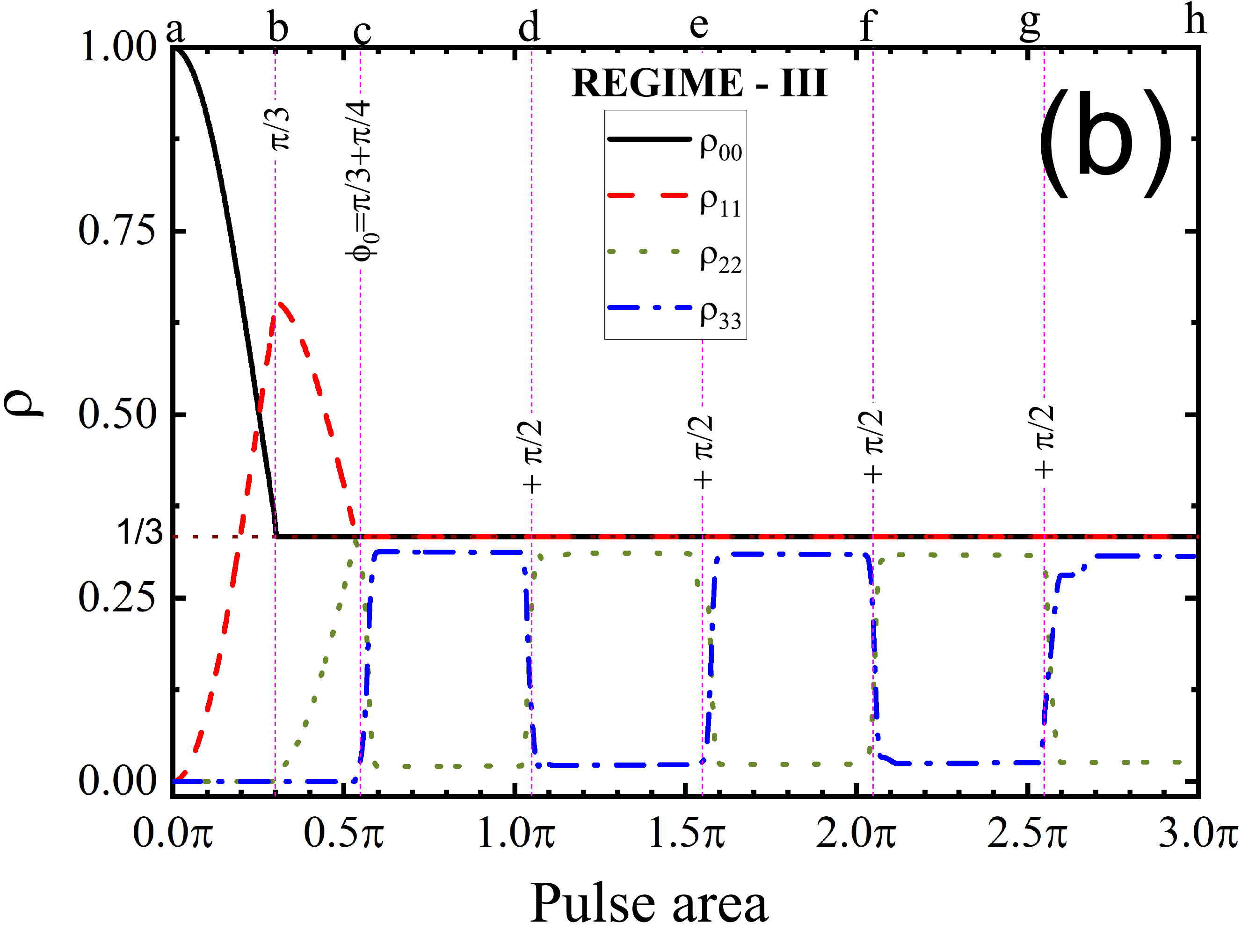}
\caption{Time evolution of population in each
state corresponding to the CNOT operation. The
figure (a) is obtained when a square pulses of
constant amplitude is shined while the figure
(b) is obtained when a pulsed laser
of width 5~fs and pulse area of $+\pi/2$ is applied
to induce CNOT operation.}\label{fig4}
\end{figure}

We now appropriately apply third pulse at time $t = {\tau}_{2}$. 
%The third pulse is chosen to be right circularly polarized pulse resonant with $|2\rangle$ $\rightleftharpoons$ $|3\rangle$ transitions. At the onset of third pulse, the Bloch vectors are $u({\tau}_{2})$, $v({\tau}_{2})$ and $w({\tau}_{2})$.
The Bloch vectors for the regime-III are found to be

\begin{subequations}
\begin{eqnarray}
u(t) &=& \frac{w''(0)}{2}(\cos(\beta_{23}t)-1)e^{-\frac{t}{T_{2}}},\\
v(t) &=& -w''(0)\sin(\beta_{23}t)e^{-\frac{t}{T_{2}}}  \\
w(t) &=&w''(0)\left[\frac{1-\cos(\beta_{23}t)}{2}-\sin(\beta_{23}t)\right]e^{-\frac{t}{T_{2}}}. \label{uvw3}
\end{eqnarray}
\end{subequations}
In obtaining these conditions, we have neglected the
role of free induction 
decay. 

For numerical analysis, we have chosen the pulse area in regime-II to be $\pi/4$ such that the total pulse area starting from the first region becomes $\pi/3 + \pi/4$. At this point of time, third pulse is applied. The diagonal elements
of $\rho$, in the Regime-III are depicted in Fig.
\ref{fig4} where we  have shown the regions of various pulse area denoted by a, b, c, d, f, g and h at the top of the figure. a denotes the start of the first regime which lasts up-to the pulse area $\pi/3$. In this region, nearly $1/3^{rd}$ of the the population remains in state $|0\rangle$ while $2/3^{rd}$ population is excited to state $|1\rangle$. At this pulse area the first pulse is switched off and the population may undergo free induction decay to state $|0\rangle$ which is neglected here. The second pulse which is switched on at $\pi/3$ pulse area now excites the remaining $2/3^{rd}$ population in state $|1\rangle$ to state $|2\rangle$.  At $\pi/4$ pulse area in the second region, half of the population $(1/2$ of $2/3)$ is raised to state $|2\rangle$ while rest of the population remains in state $|1\rangle$. At this juncture, third pulse is applied  which gives rise to stimulated de excitation of population from state $|2\rangle$ to state $|3\rangle$. The states $|0\rangle$ and $|1\rangle$
are freeze and are allowed to dephase spontaneously.
However, for time smaller than the dephasing time, the populations in states $|0\rangle$ and $|1\rangle$ nearly
remains constant as shown in the figure \ref{fig4},
while the states $|2\rangle$  and $|3\rangle$ are
allowed to execute optical nutation. It is exciting
to note that the application of every $\pi/2$ pulse, population
flips/flops between the states $|2\rangle\rightleftharpoons
|3\rangle$. 

Now the system
is said to be prepared for the demonstration of
quantum\ CNOT\ operation. The total pulse area applied
to prepare the system is defined as $\phi_0=\frac{\pi}{3}+\frac{\pi}{4}$.  

\subsection{CNOT Operation}

To enable us to understand the CNOT operation, 
we shall redefine the states from decimal coding
to binary coding, discussed earlier as $|0\rangle \rightleftharpoons
|00\rangle$, $|1\rangle \rightleftharpoons
|01\rangle$, $|2\rangle \rightleftharpoons
|10\rangle$ and $|3\rangle \rightleftharpoons
|11\rangle$. The matrix representation of the
states are
\begin{eqnarray}
|00\rangle&=\begin{bmatrix}1&0&0&0\\ 0&0&0&0\\0&0&0&0\\0&0&0&0\end{bmatrix}\hspace{-2mm},
|01\rangle&=\begin{bmatrix}0&0&0&0\\ 0&1&0&0\\0&0&0&0\\0&0&0&0\end{bmatrix}\hspace{-2mm},\nonumber\\
|10\rangle&=\begin{bmatrix}0&0&0&0\\ 0&0&0&0\\0&0&1&0\\0&0&0&0\end{bmatrix}\hspace{-2mm},
|11\rangle&=\begin{bmatrix}0&0&0&0\\ 0&0&0&0\\0&0&0&0\\0&0&0&1\end{bmatrix}
\end{eqnarray}
At $t=\tau_2$, the input states
defined as, ${|\psi_{in}\rangle}=|00\rangle + |01\rangle + |10\rangle + |11\rangle$
\begin{equation}
{|\psi_{in}\rangle}=\begin{bmatrix} 1&0&0&0\\ 0&1&0&0\\0&0&1&0\\0&0&0&1 \end{bmatrix}\hspace{-2mm}.
\end{equation}
Application of $\pi/2$ pulse flips the $|10\rangle\rightarrow|11\rangle$
and $|11\rangle\rightarrow|10\rangle$. Hence,
the output state shall be defined as
\begin{equation}
{|\psi_{out}\rangle}=\begin{bmatrix} 1&0&0&0\\ 0&1&0&0\\0&0&0&1\\0&0&1&0 \end{bmatrix}.
\end{equation}
In other words, the unitary matrix corresponding to these transformation
is $U_{CNOT}$. The general picture of CNOT operation can be realised in the recipe 
described in the above formulation. In order to obtain $|\psi_{out}\rangle$, we choose
proper pulse area in pulse sequencing. 
In practical situation, neither the excitation pulse is ideally 
monochromatic nor one can select a system of identical quantum dots. 
Under such circumstance, the recipe described above gives us a 
statistical picture. Figure  \ref{fig4}, indicates that the population corresponding to $|00\rangle$ and $|01\rangle$ states remains constant at $1/3^{rd}$ while the remaining $1/3^{rd}$ population at levels $|10\rangle$ and $|11\rangle$ undergoes desired switching.

\begin{figure}
    \centering
 \includegraphics[width=.23\columnwidth]{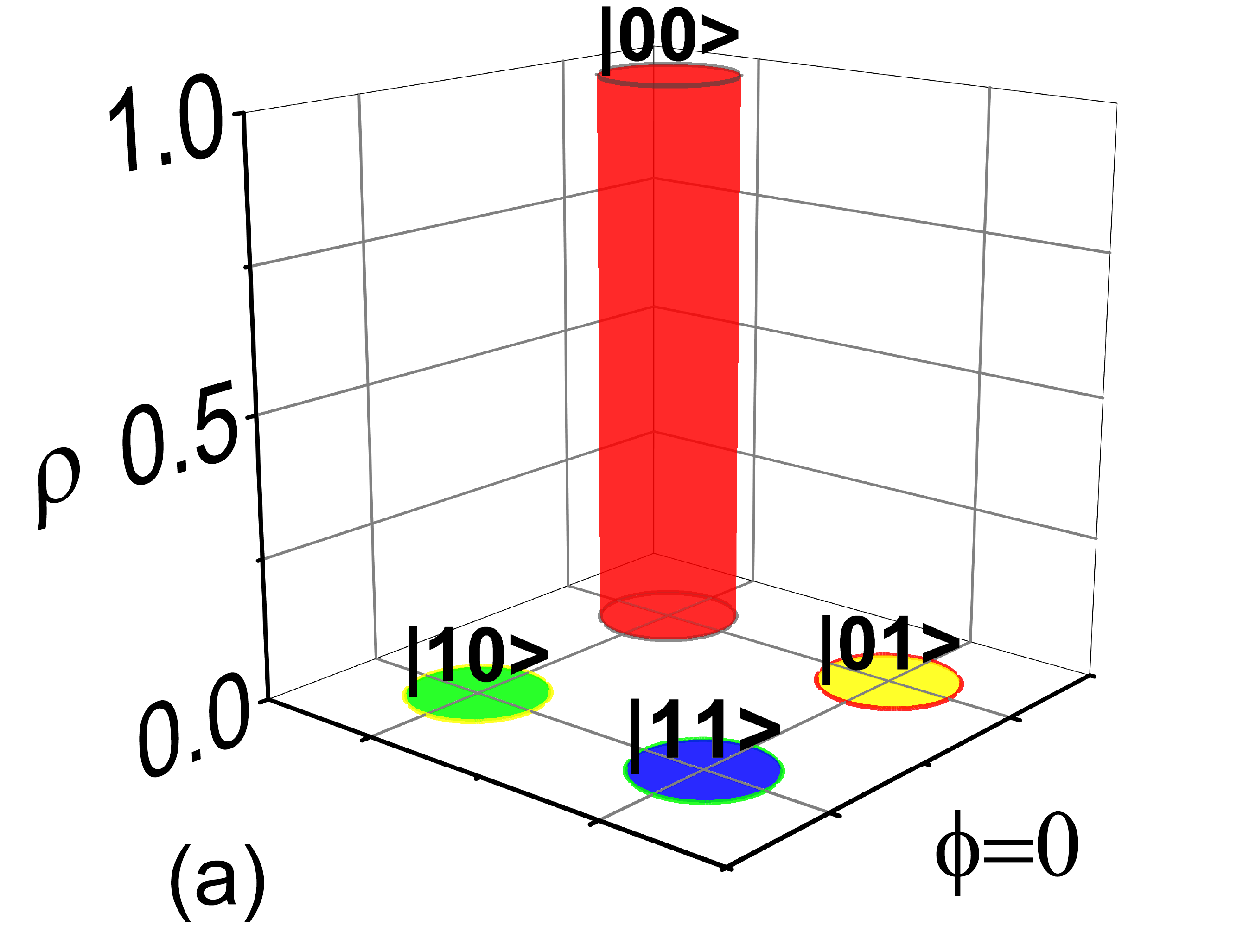}
  \includegraphics[width=.23\columnwidth]{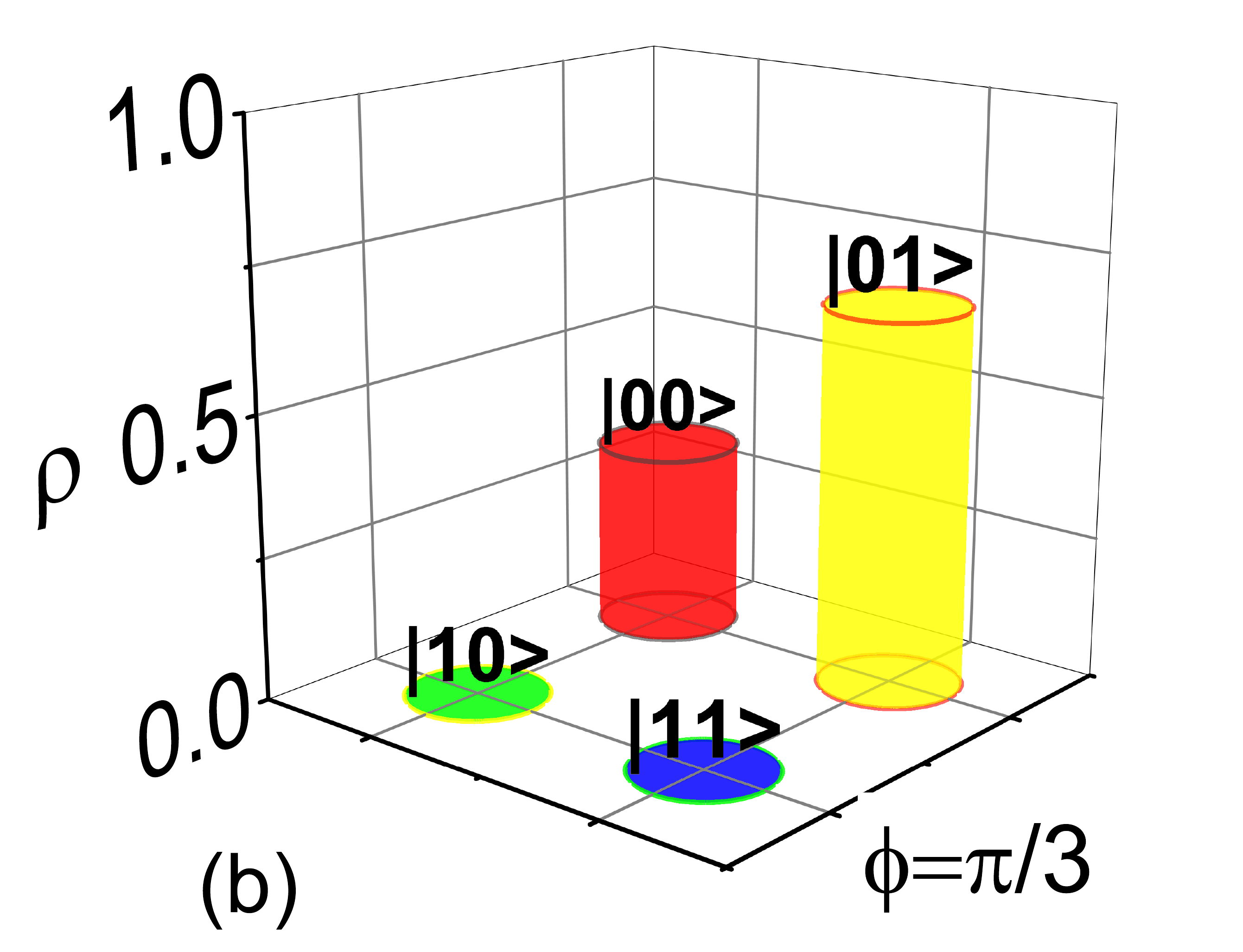}
   \includegraphics[width=.230\columnwidth]{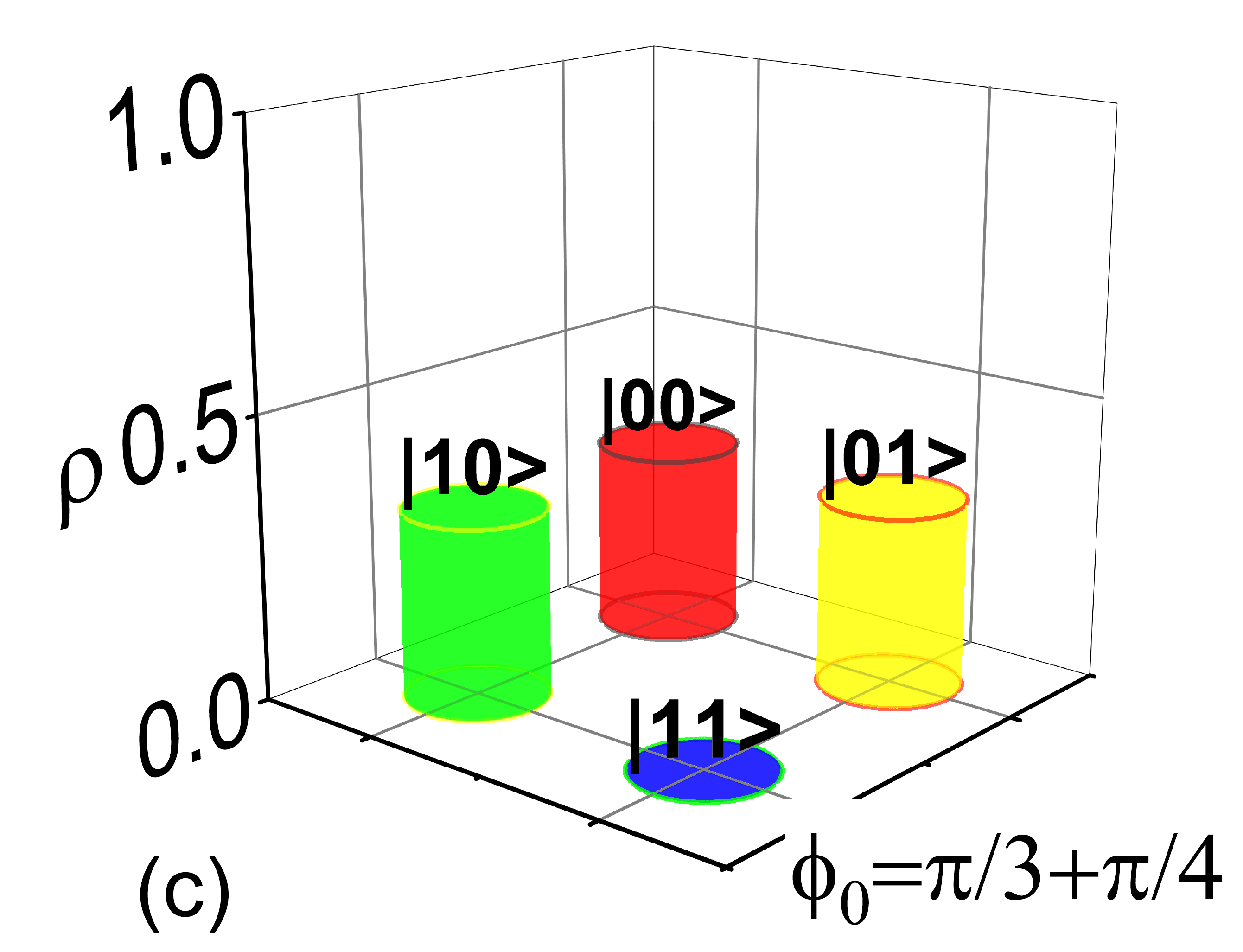}
    \includegraphics[width=.230\columnwidth]{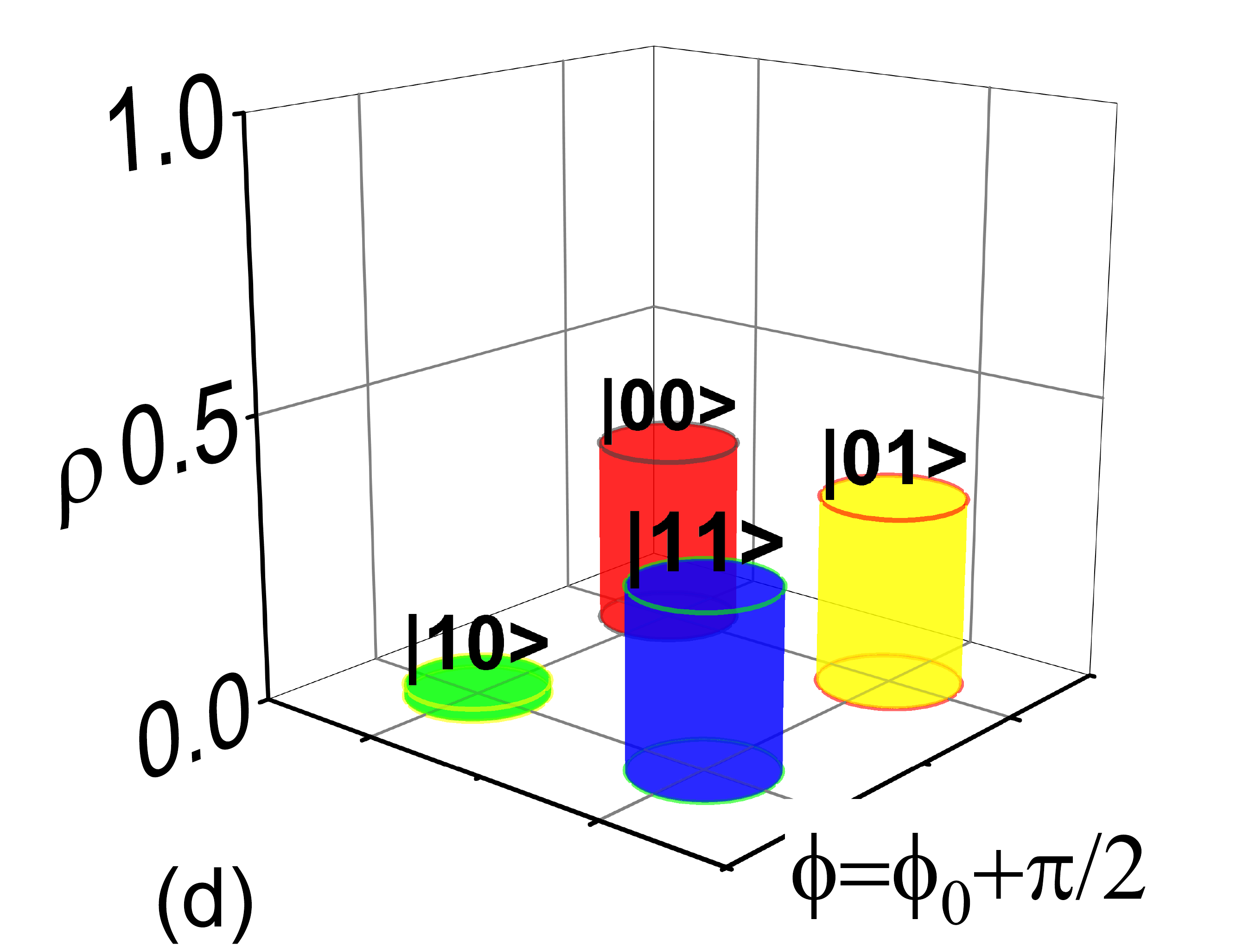}\\
 \includegraphics[width=.23\columnwidth]{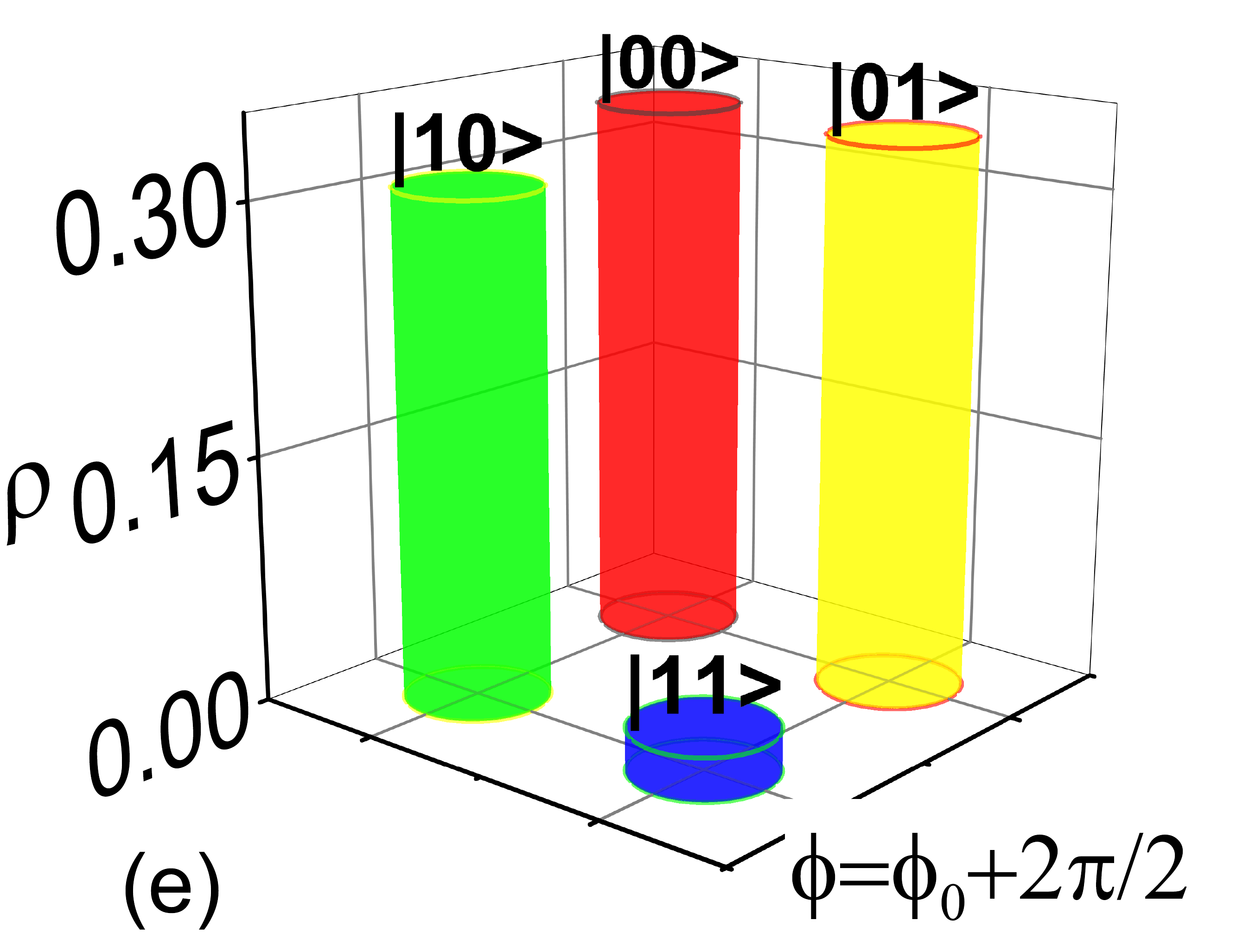}
  \includegraphics[width=.23\columnwidth]{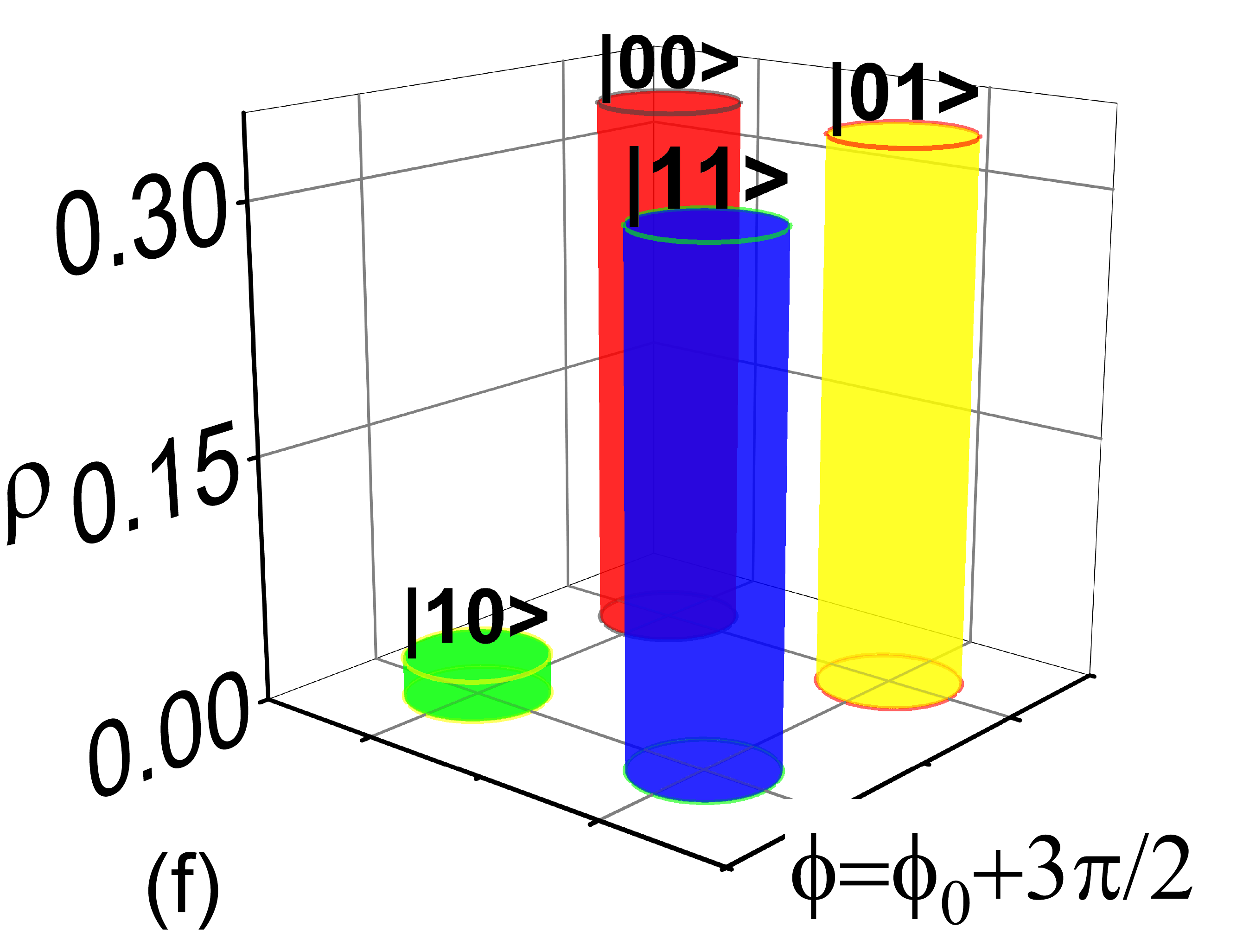}
   \includegraphics[width=.23\columnwidth]{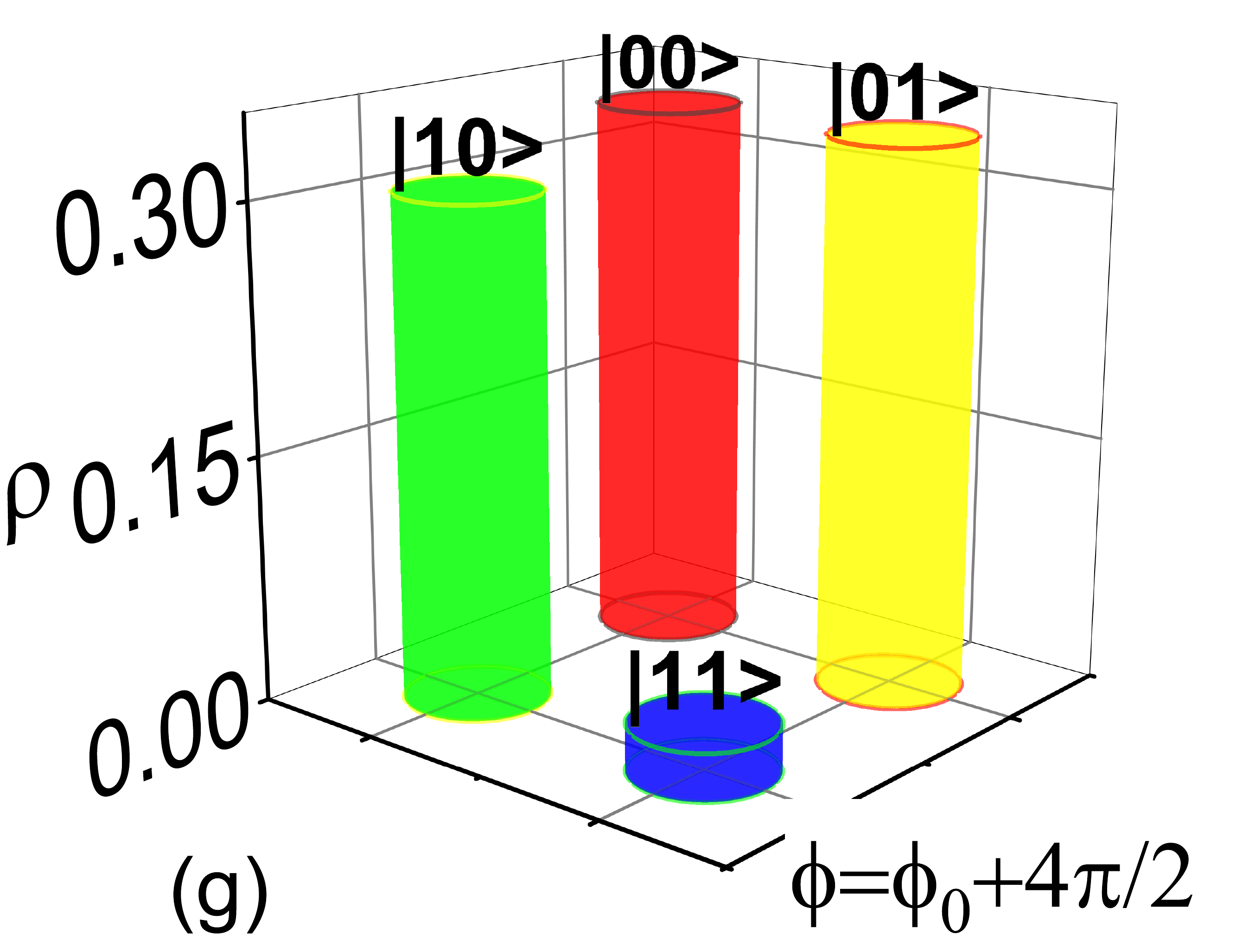}
    \includegraphics[width=.23\columnwidth]{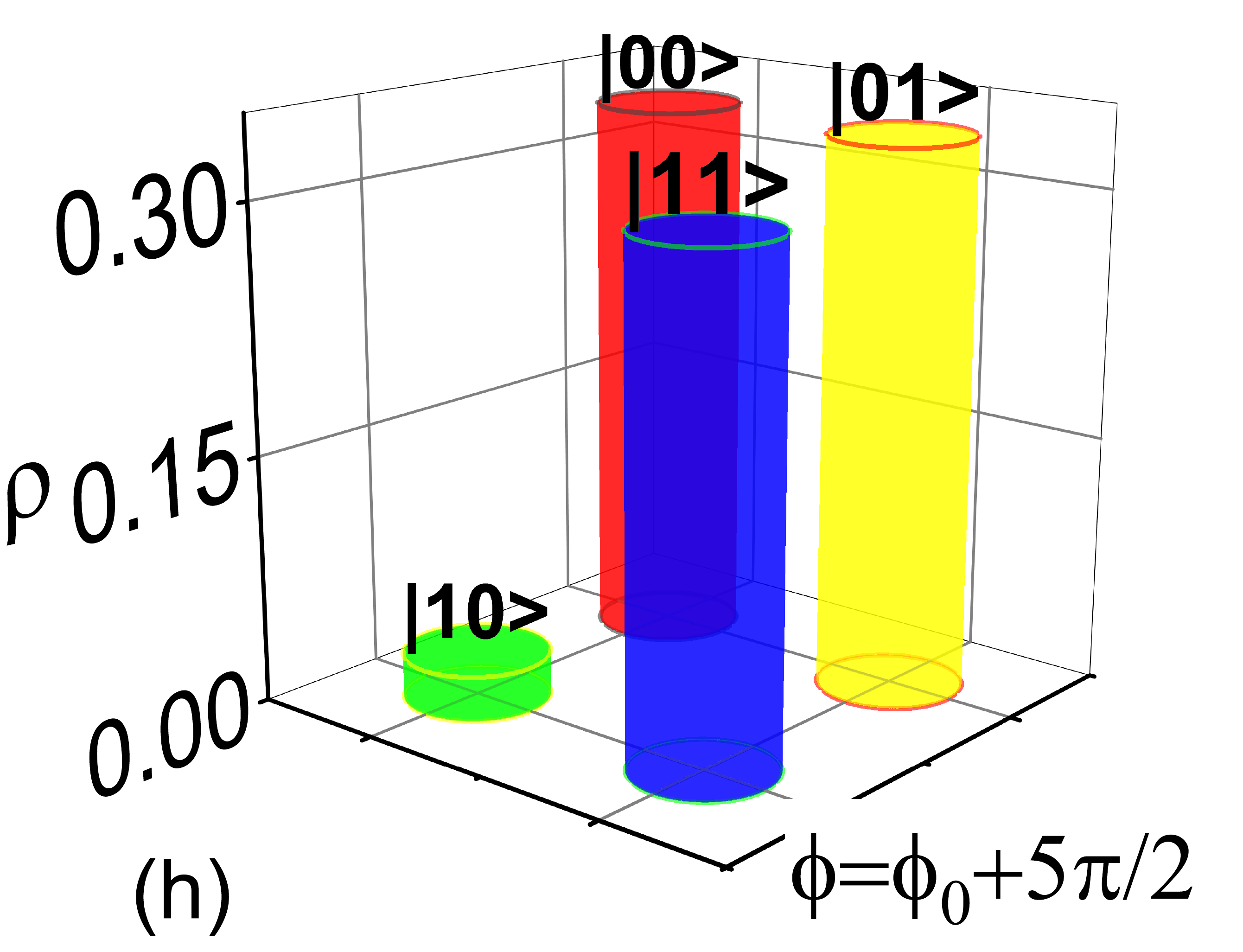} \\
 \includegraphics[width=.1\columnwidth,height=.1\columnwidth]{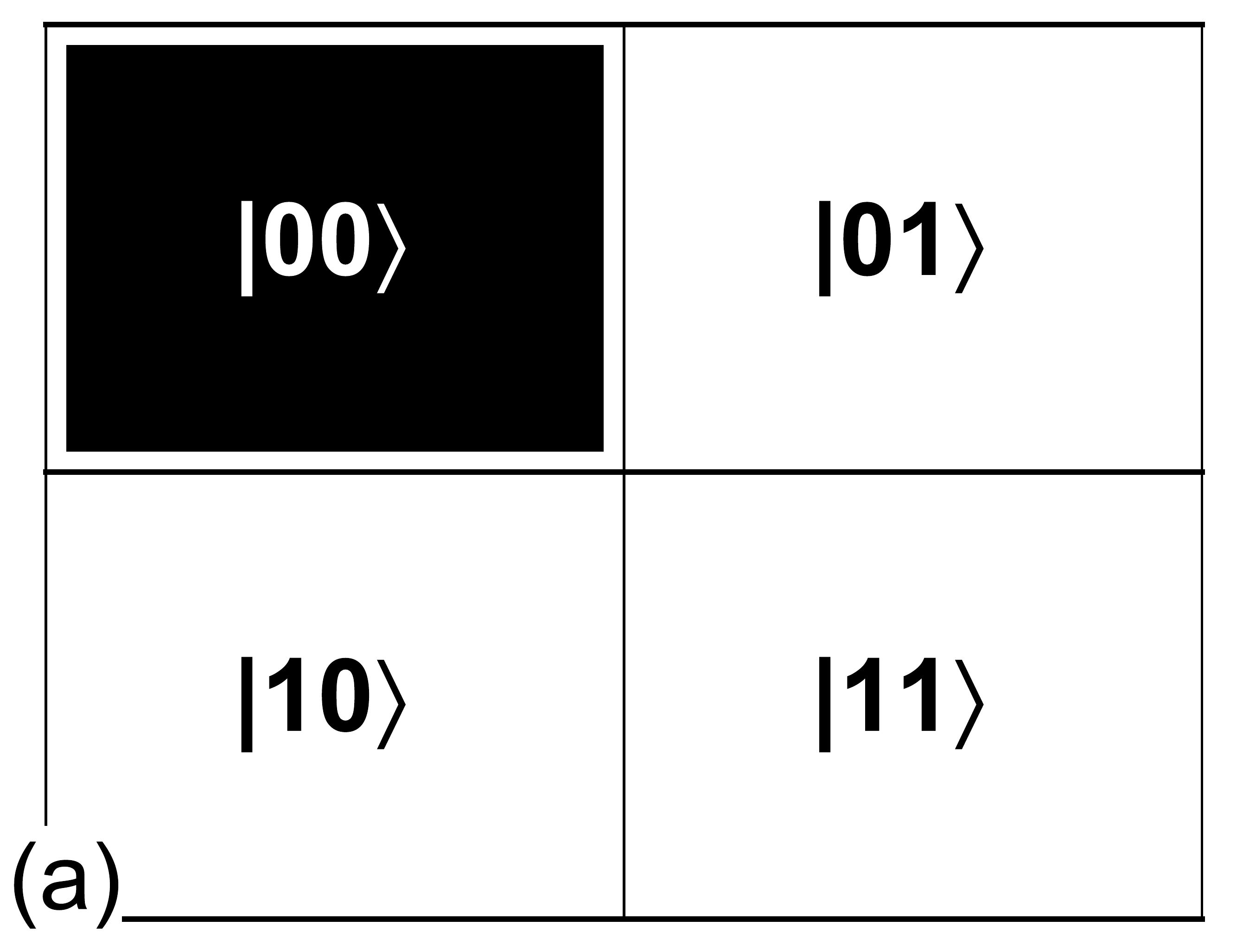}
 \includegraphics[width=.1\columnwidth,height=.1\columnwidth]{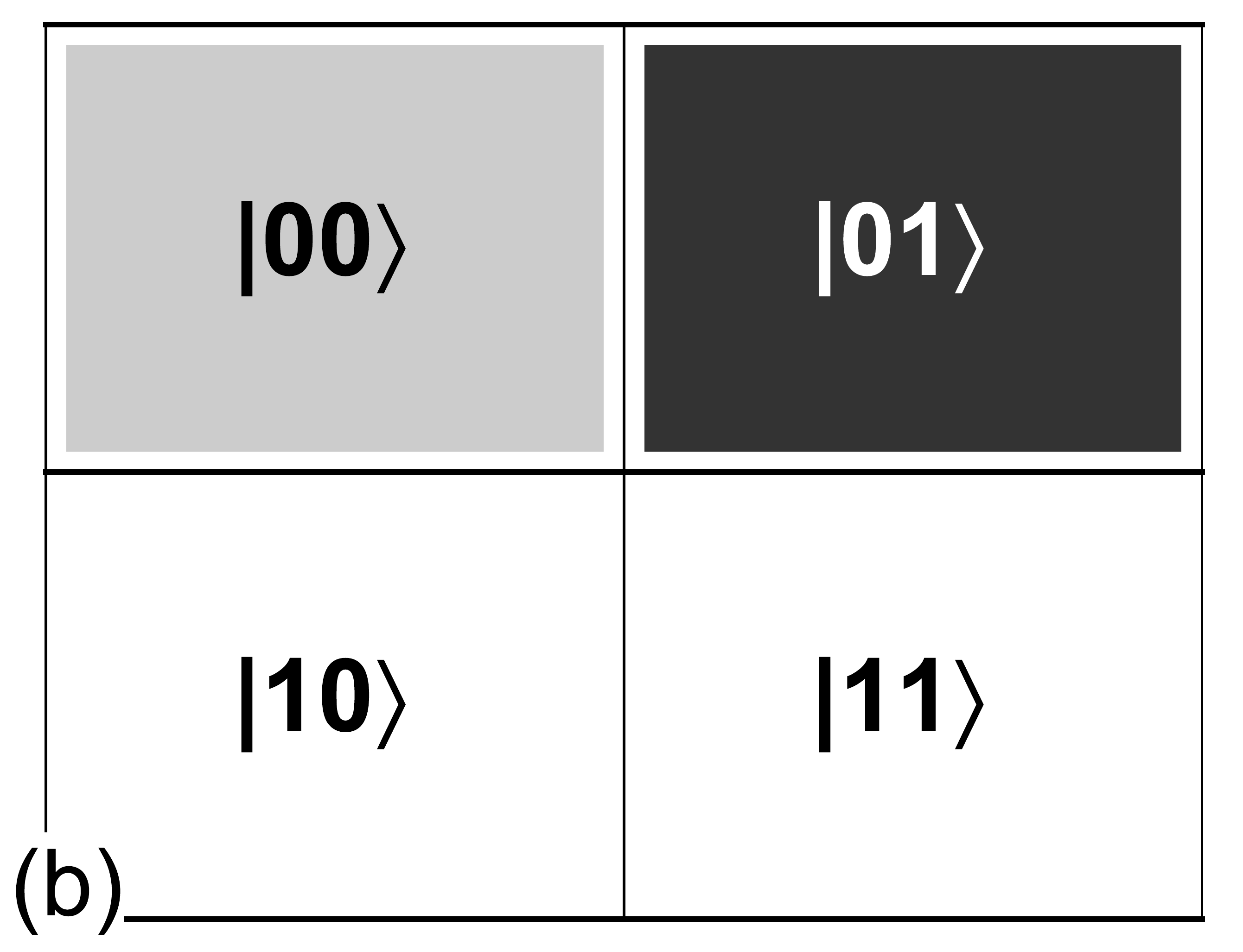}
 \includegraphics[width=.1\columnwidth,height=.1\columnwidth]{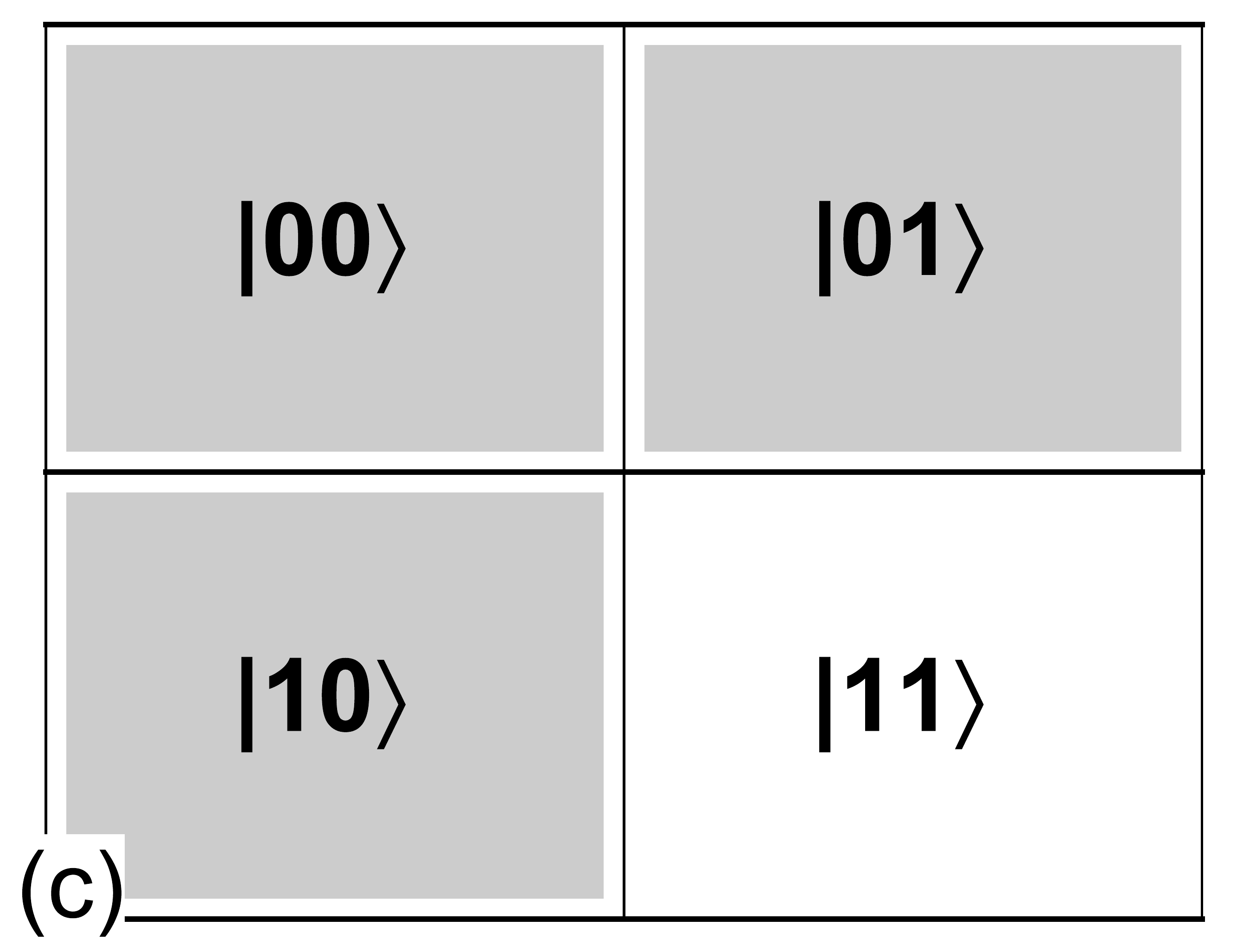}
 \includegraphics[width=.1\columnwidth,height=.1\columnwidth]{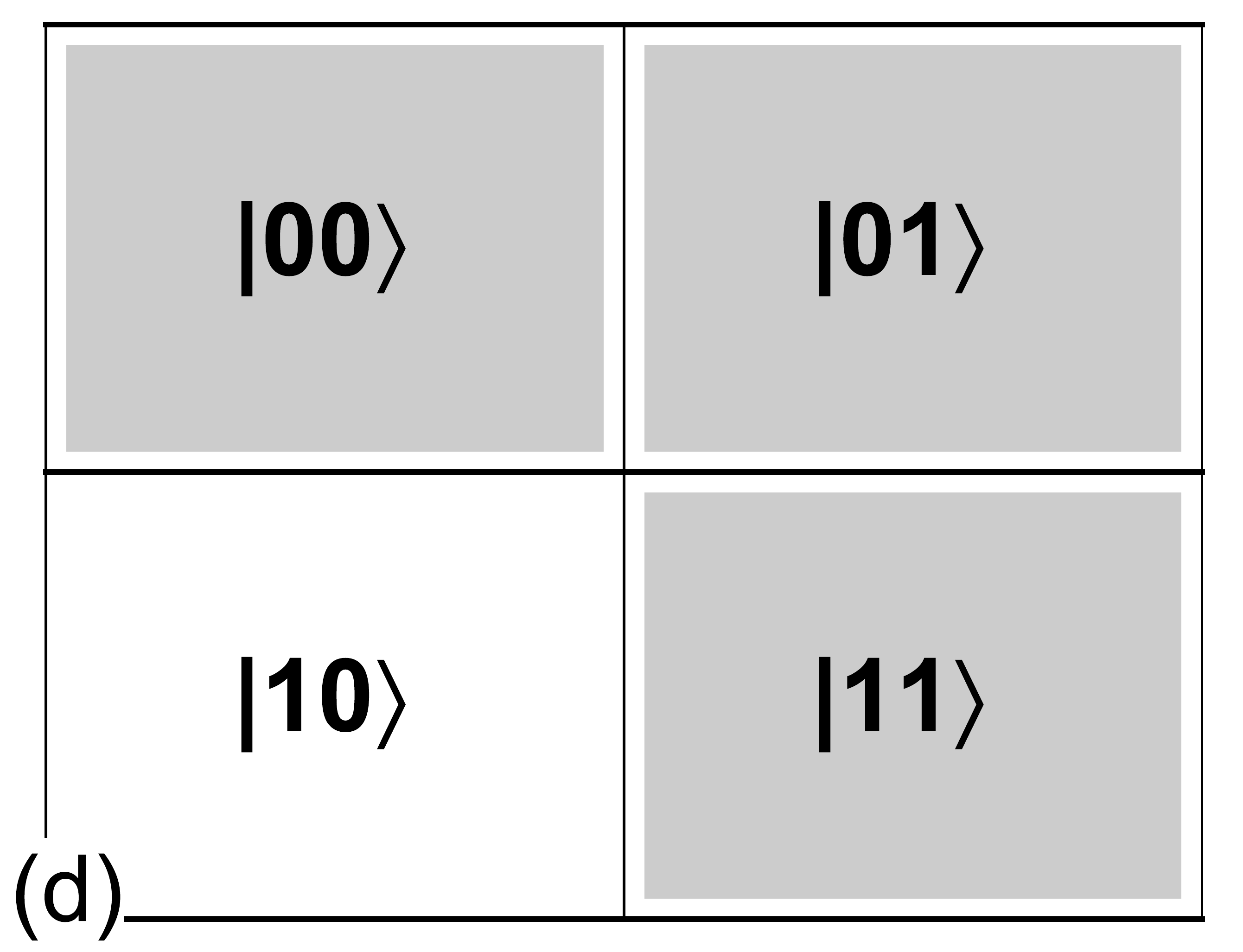}
 \includegraphics[width=.1\columnwidth,height=.1\columnwidth]{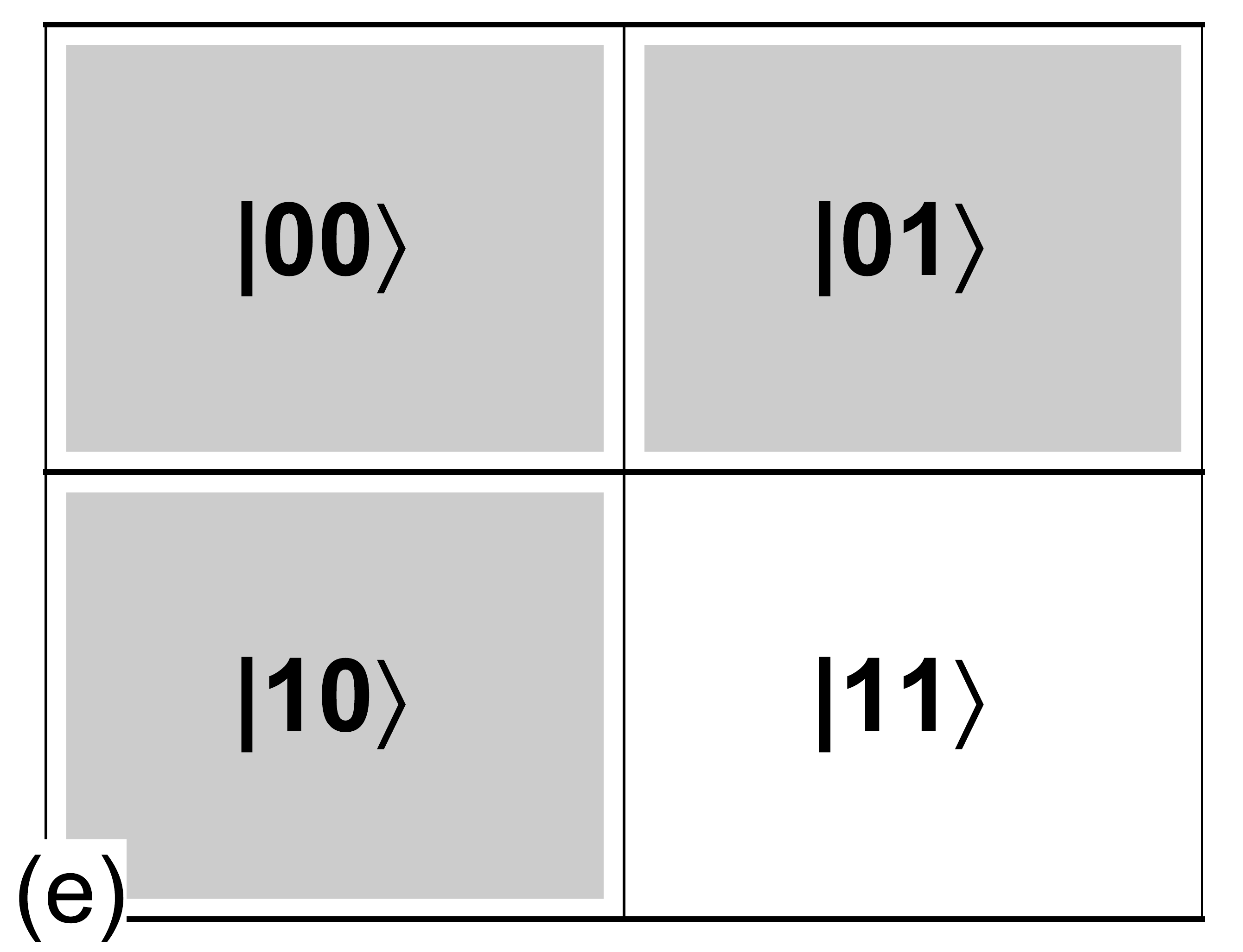}
 \includegraphics[width=.1\columnwidth,height=.1\columnwidth]{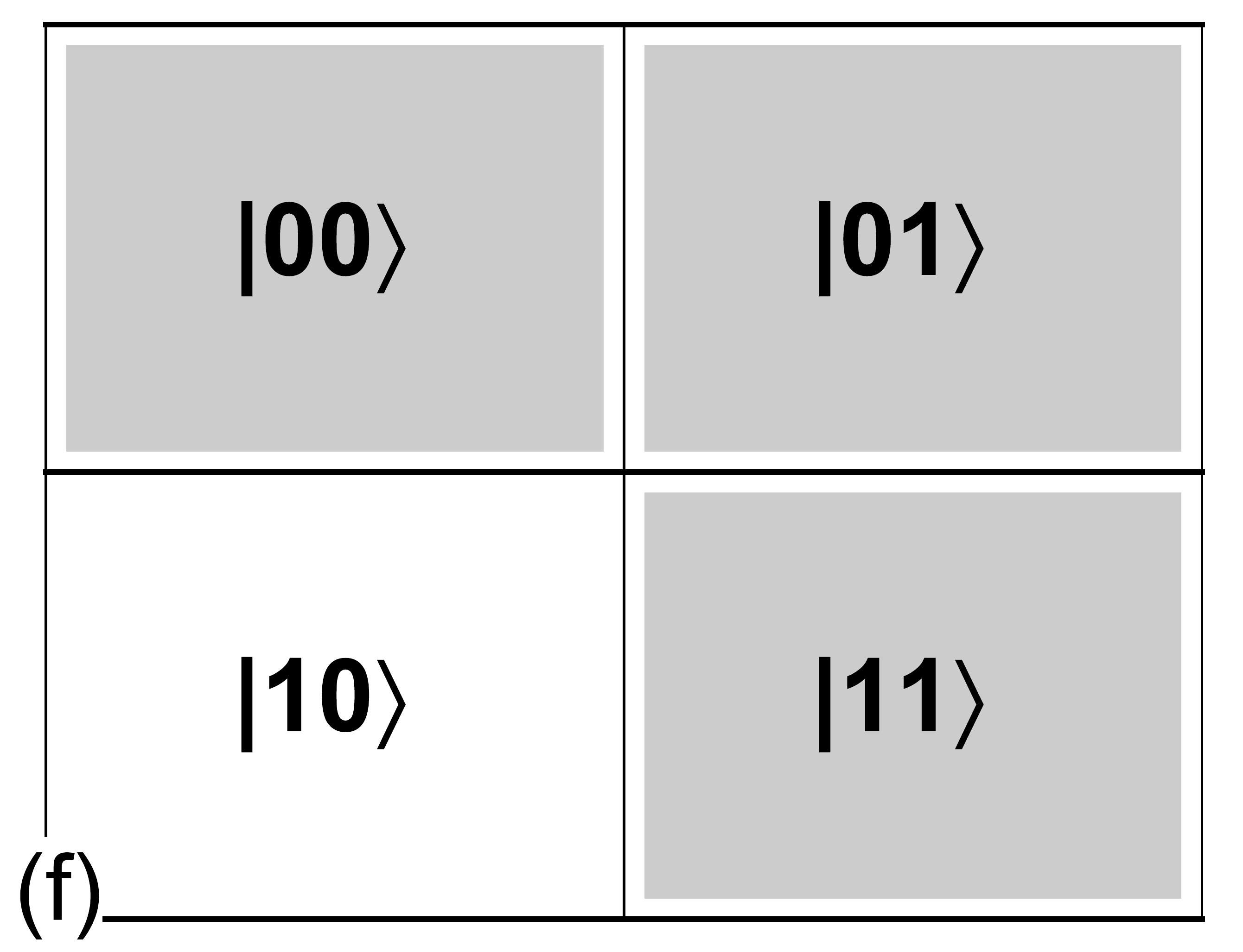}
 \includegraphics[width=.1\columnwidth,height=.1\columnwidth]{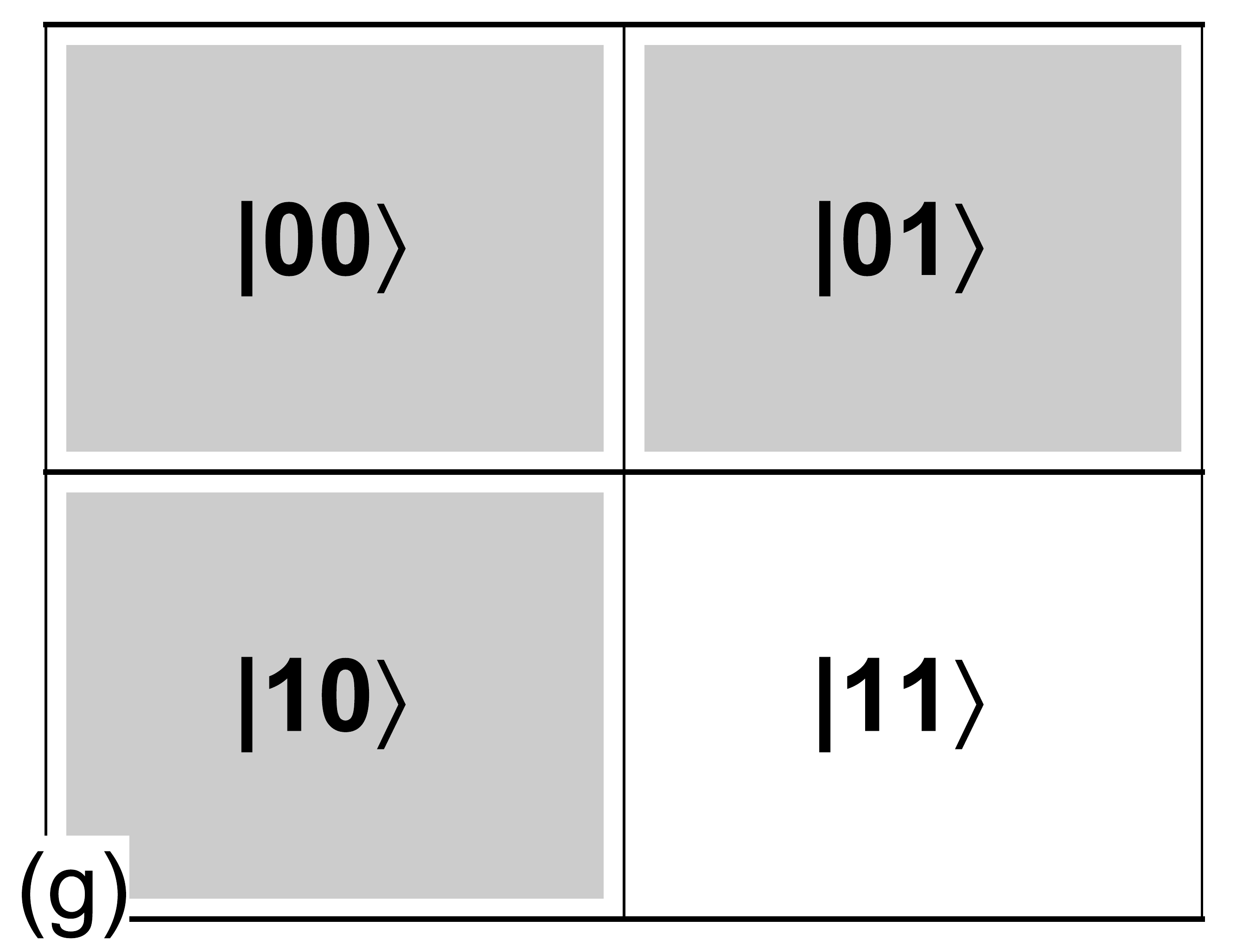}
 \includegraphics[width=.12\columnwidth,height=.1\columnwidth]{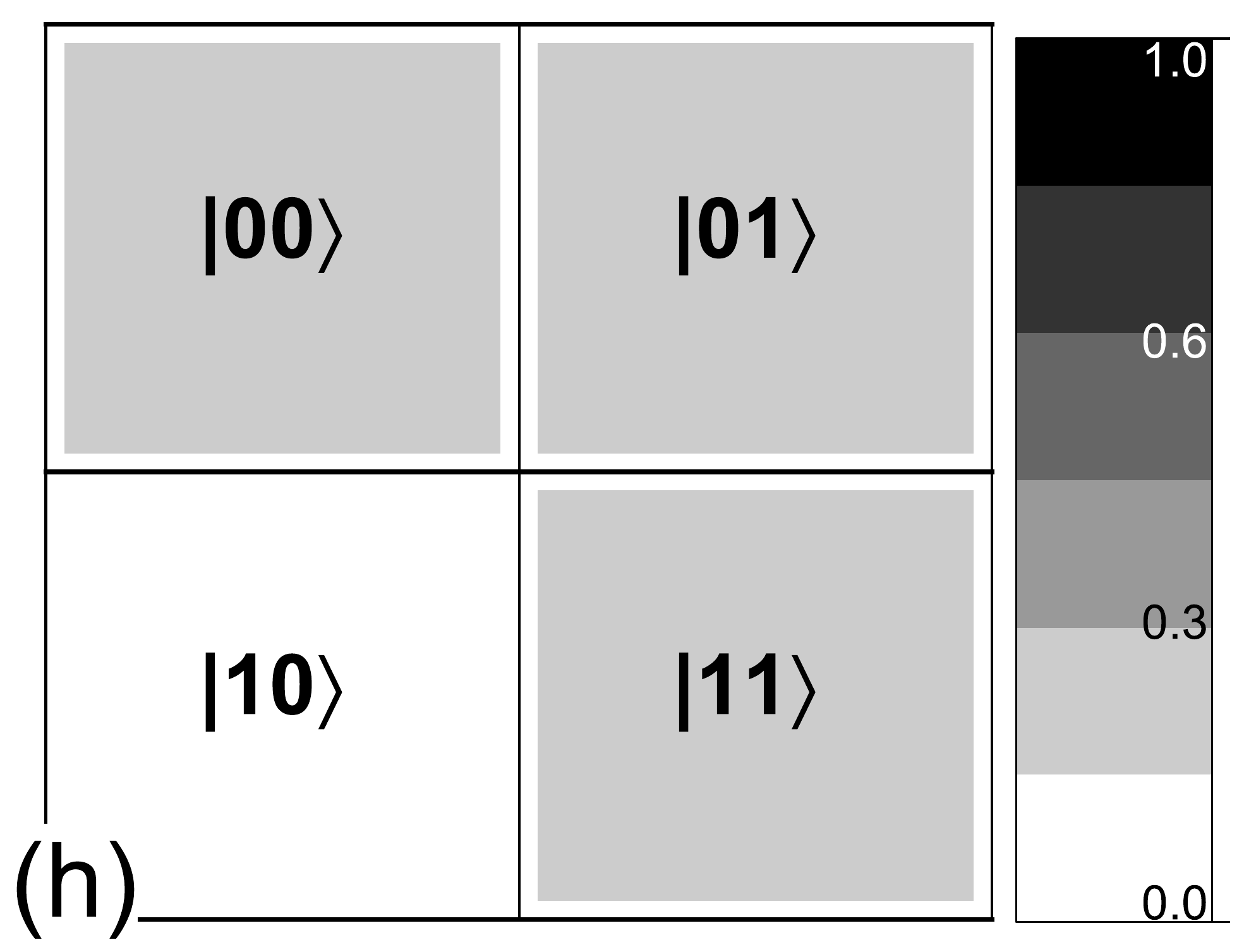}
  \caption{The 3D-bar chart and tomogram of CNOT gate demonstrated.
    The figures a-h are obtained for the pulse
    area as shown in Figures 5A and 5B.}\label{fig6}
\end{figure}

\begin{figure}
\includegraphics[width=.85\columnwidth]{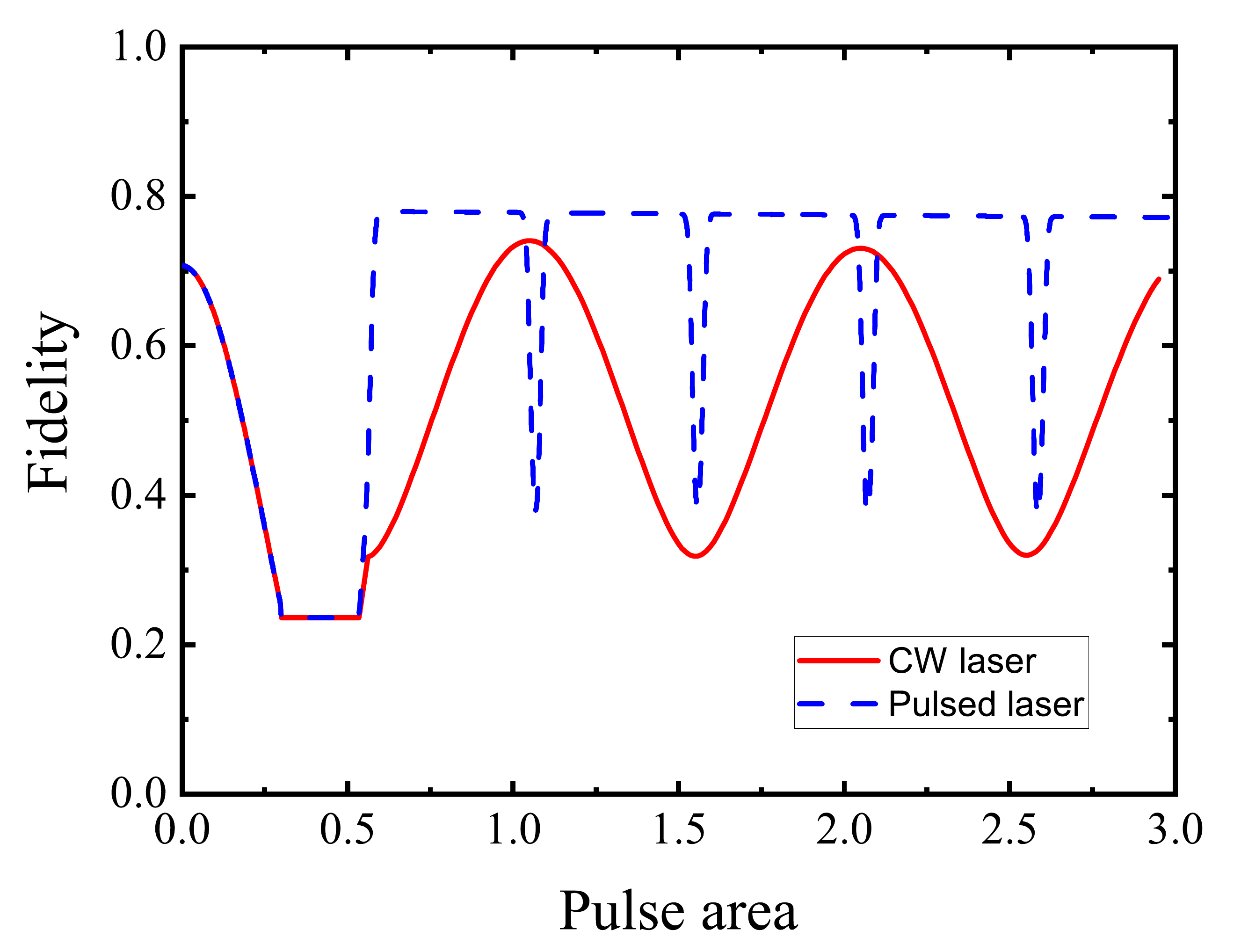}
\caption{The measured fidelity obtained for the cases discussed in Figs. 5A and 5B.}\label{Fig7}
\end{figure} 

A clearer picture of the same can be seen in figure \ref{fig6}, where we have demonstrated the populations in each state for every
$\pi/2$ pulse area, ranging from $\phi=\phi_0$ to $\phi=\phi_0+\frac{5\pi}{2}$.
Figures \ref{fig6}a - \ref{fig6}h are obtained for area of the excitation  pulses corresponding to the locations
a-h as shown in Fig. \ref{fig4}. The respective tomograms are shown at the bottom of the same figure. The initialization of the CNOT gates is achieved when a pulse area of $\phi_0=\frac{\pi}{3}+\frac{\pi}{4}$ is applied (fig. \ref{fig6}c). One can observe from the figure that flipping between $|10\rangle$ and $|11\rangle$ state occurs at pulse area $\pi/2$. However, demonstration of an efficient CNOT gate is incomplete without mentioning its fidelity. In the present formulation we create a Bell state by combining  CNOT with single qubits. 

From the above figures (\ref{fig6}d, \ref{fig6}f, \ref{fig6}h) it is clear that for input states $(|00\rangle, |01\rangle, |10\rangle, |11\rangle)$ the target states are Bell state, which are obtained when a pulse area  of $\pi$ is applied. On the other hand the target states corresponding to figure (\ref{fig6}c, \ref{fig6}e, \ref{fig6}g) are non Bell states.

We now address ourselves to the fidelity of the CNOT operation. The fidelity of the prepared states in this two qubit operation is given by $\sqrt{\langle \Psi_{target}|\rho|\Psi_{target}\rangle}$, where $\Psi_{target}$ is the Bell state given by $\frac{1}{\sqrt{2}}(|00\rangle-i|11\rangle)$ \cite{zajac2018resonantly, bengtsson2017geometry}. In figure \ref{Fig7}, we have exhibited the fidelity as a function of pulse area
for a square pulse as well as for the Gaussian
pulse. One may notice that the maximum fidelity obtainable after the initialization is $33\%$ at pulse area $\pi/3+\pi/4$. We further
identify that the the Bell states corresponding
to figures \ref{fig6}d, \ref{fig6}f and \ref{fig6}h yields
maximum fidelity of 74\% with square pulses while
with Gaussian pulsed lasers a maximum fidelity
of 80\% is predicted.  
\section{CONCLUSIONS}
In conclusion, we have studied  the evolution
of population in heavy-hole valance band and conduction
band states
in a magnetic impurity doped semiconductor
quantum dots. The conduction band states are entangled
via a microwave wave pulse. In this system series
of pulses of pre-decided pulse area are         chosen
to excite the population from the heavy-hole state $|+\frac{3}{2}\rangle$
to valance band state $|-\frac{3}{2}\rangle$.
This sequencing of pulses allows us to prepare
the system for  CNOT gate operation. We have also
calculated the fidelity of CNOT\ gate and found
to have a maximum value of $\approx 80$\%.  
\section{Acknowledgment}
The authors thank Prof. P. K. Sen for 
discussions. PS thank SERB-DST and University Grants Commission, New Delhi for
financial support. JTA acknowledges the financial
support received from RPS-AICTE, New Delhi and SERB-DST,
New Delhi. 
\bibliographystyle{apsrev4-1}
\bibliography{References} 

\begin{thebibliography}{29}%
\makeatletter
\providecommand \@ifxundefined [1]{%
 \@ifx{#1\undefined}
}%
\providecommand \@ifnum [1]{%
 \ifnum #1\expandafter \@firstoftwo
 \else \expandafter \@secondoftwo
 \fi
}%
\providecommand \@ifx [1]{%
 \ifx #1\expandafter \@firstoftwo
 \else \expandafter \@secondoftwo
 \fi
}%
\providecommand \natexlab [1]{#1}%
\providecommand \enquote  [1]{``#1''}%
\providecommand \bibnamefont  [1]{#1}%
\providecommand \bibfnamefont [1]{#1}%
\providecommand \citenamefont [1]{#1}%
\providecommand \href@noop [0]{\@secondoftwo}%
\providecommand \href [0]{\begingroup \@sanitize@url \@href}%
\providecommand \@href[1]{\@@startlink{#1}\@@href}%
\providecommand \@@href[1]{\endgroup#1\@@endlink}%
\providecommand \@sanitize@url [0]{\catcode `\\12\catcode `\$12\catcode
  `\&12\catcode `\#12\catcode `\^12\catcode `\_12\catcode `\%12\relax}%
\providecommand \@@startlink[1]{}%
\providecommand \@@endlink[0]{}%
\providecommand \url  [0]{\begingroup\@sanitize@url \@url }%
\providecommand \@url [1]{\endgroup\@href {#1}{\urlprefix }}%
\providecommand \urlprefix  [0]{URL }%
\providecommand \Eprint [0]{\href }%
\providecommand \doibase [0]{http://dx.doi.org/}%
\providecommand \selectlanguage [0]{\@gobble}%
\providecommand \bibinfo  [0]{\@secondoftwo}%
\providecommand \bibfield  [0]{\@secondoftwo}%
\providecommand \translation [1]{[#1]}%
\providecommand \BibitemOpen [0]{}%
\providecommand \bibitemStop [0]{}%
\providecommand \bibitemNoStop [0]{.\EOS\space}%
\providecommand \EOS [0]{\spacefactor3000\relax}%
\providecommand \BibitemShut  [1]{\csname bibitem#1\endcsname}%
\let\auto@bib@innerbib\@empty
%</preamble>
\bibitem [{\citenamefont {Loss}\ and\ \citenamefont
  {DiVincenzo}(1998)}]{loss1998quantum}%
  \BibitemOpen
  \bibfield  {author} {\bibinfo {author} {\bibfnamefont {D.}~\bibnamefont
  {Loss}}\ and\ \bibinfo {author} {\bibfnamefont {D.~P.}\ \bibnamefont
  {DiVincenzo}},\ }\href@noop {} {\bibfield  {journal} {\bibinfo  {journal}
  {Physical Review A}\ }\textbf {\bibinfo {volume} {57}},\ \bibinfo {pages}
  {120} (\bibinfo {year} {1998})}\BibitemShut {NoStop}%
\bibitem [{\citenamefont {DiVincenzo}(2000)}]{divincenzo2000physical}%
  \BibitemOpen
  \bibfield  {author} {\bibinfo {author} {\bibfnamefont {D.~P.}\ \bibnamefont
  {DiVincenzo}},\ }\href@noop {} {\bibfield  {journal} {\bibinfo  {journal}
  {Fortschritte der Physik: Progress of Physics}\ }\textbf {\bibinfo {volume}
  {48}},\ \bibinfo {pages} {771} (\bibinfo {year} {2000})}\BibitemShut
  {NoStop}%
\bibitem [{\citenamefont {Puri}\ \emph {et~al.}(2017)\citenamefont {Puri},
  \citenamefont {McMahon},\ and\ \citenamefont {Yamamoto}}]{puri2017universal}%
  \BibitemOpen
  \bibfield  {author} {\bibinfo {author} {\bibfnamefont {S.}~\bibnamefont
  {Puri}}, \bibinfo {author} {\bibfnamefont {P.~L.}\ \bibnamefont {McMahon}}, \
  and\ \bibinfo {author} {\bibfnamefont {Y.}~\bibnamefont {Yamamoto}},\
  }\href@noop {} {\bibfield  {journal} {\bibinfo  {journal} {Physical Review
  B}\ }\textbf {\bibinfo {volume} {95}},\ \bibinfo {pages} {125410} (\bibinfo
  {year} {2017})}\BibitemShut {NoStop}%
\bibitem [{\citenamefont {Mills}\ \emph {et~al.}(2019)\citenamefont {Mills},
  \citenamefont {Zajac}, \citenamefont {Gullans}, \citenamefont {Schupp},
  \citenamefont {Hazard},\ and\ \citenamefont {Petta}}]{mills2019shuttling}%
  \BibitemOpen
  \bibfield  {author} {\bibinfo {author} {\bibfnamefont {A.}~\bibnamefont
  {Mills}}, \bibinfo {author} {\bibfnamefont {D.}~\bibnamefont {Zajac}},
  \bibinfo {author} {\bibfnamefont {M.}~\bibnamefont {Gullans}}, \bibinfo
  {author} {\bibfnamefont {F.}~\bibnamefont {Schupp}}, \bibinfo {author}
  {\bibfnamefont {T.}~\bibnamefont {Hazard}}, \ and\ \bibinfo {author}
  {\bibfnamefont {J.~R.}\ \bibnamefont {Petta}},\ }\href@noop {} {\bibfield
  {journal} {\bibinfo  {journal} {Nature communications}\ }\textbf {\bibinfo
  {volume} {10}},\ \bibinfo {pages} {1063} (\bibinfo {year}
  {2019})}\BibitemShut {NoStop}%
\bibitem [{\citenamefont {Wenz}\ \emph {et~al.}(2019)\citenamefont {Wenz},
  \citenamefont {Klochan}, \citenamefont {Hohls}, \citenamefont {Gerster},
  \citenamefont {Kashcheyevs},\ and\ \citenamefont
  {Schumacher}}]{wenz2019quantum}%
  \BibitemOpen
  \bibfield  {author} {\bibinfo {author} {\bibfnamefont {T.}~\bibnamefont
  {Wenz}}, \bibinfo {author} {\bibfnamefont {J.}~\bibnamefont {Klochan}},
  \bibinfo {author} {\bibfnamefont {F.}~\bibnamefont {Hohls}}, \bibinfo
  {author} {\bibfnamefont {T.}~\bibnamefont {Gerster}}, \bibinfo {author}
  {\bibfnamefont {V.}~\bibnamefont {Kashcheyevs}}, \ and\ \bibinfo {author}
  {\bibfnamefont {H.~W.}\ \bibnamefont {Schumacher}},\ }\href@noop {}
  {\bibfield  {journal} {\bibinfo  {journal} {Physical Review B}\ }\textbf
  {\bibinfo {volume} {99}},\ \bibinfo {pages} {201409} (\bibinfo {year}
  {2019})}\BibitemShut {NoStop}%
\bibitem [{\citenamefont {Qureshi}\ \emph {et~al.}(2008)\citenamefont
  {Qureshi}, \citenamefont {Sen}, \citenamefont {Andrews},\ and\ \citenamefont
  {Sen}}]{qureshi2008all}%
  \BibitemOpen
  \bibfield  {author} {\bibinfo {author} {\bibfnamefont {M.~S.}\ \bibnamefont
  {Qureshi}}, \bibinfo {author} {\bibfnamefont {P.}~\bibnamefont {Sen}},
  \bibinfo {author} {\bibfnamefont {J.}~\bibnamefont {Andrews}}, \ and\
  \bibinfo {author} {\bibfnamefont {P.~K.}\ \bibnamefont {Sen}},\ }\href@noop
  {} {\bibfield  {journal} {\bibinfo  {journal} {IEEE Journal of Quantum
  Electronics}\ }\textbf {\bibinfo {volume} {45}},\ \bibinfo {pages} {59}
  (\bibinfo {year} {2008})}\BibitemShut {NoStop}%
\bibitem [{\citenamefont {Awschalom}\ \emph {et~al.}(2013)\citenamefont
  {Awschalom}, \citenamefont {Loss},\ and\ \citenamefont
  {Samarth}}]{awschalom2013semiconductor}%
  \BibitemOpen
  \bibfield  {author} {\bibinfo {author} {\bibfnamefont {D.~D.}\ \bibnamefont
  {Awschalom}}, \bibinfo {author} {\bibfnamefont {D.}~\bibnamefont {Loss}}, \
  and\ \bibinfo {author} {\bibfnamefont {N.}~\bibnamefont {Samarth}},\
  }\href@noop {} {\emph {\bibinfo {title} {Semiconductor spintronics and
  quantum computation}}}\ (\bibinfo  {publisher} {Springer Science \& Business
  Media},\ \bibinfo {year} {2013})\ pp.\ \bibinfo {pages} {94--95}\BibitemShut
  {NoStop}%
\bibitem [{\citenamefont {Hanson}\ \emph {et~al.}(2004)\citenamefont {Hanson},
  \citenamefont {Elzerman}, \citenamefont {van Beveren}, \citenamefont
  {Vandersypen},\ and\ \citenamefont {Kouwenhoven}}]{hanson2004electron}%
  \BibitemOpen
  \bibfield  {author} {\bibinfo {author} {\bibfnamefont {R.}~\bibnamefont
  {Hanson}}, \bibinfo {author} {\bibfnamefont {J.}~\bibnamefont {Elzerman}},
  \bibinfo {author} {\bibfnamefont {L.~W.}\ \bibnamefont {van Beveren}},
  \bibinfo {author} {\bibfnamefont {L.}~\bibnamefont {Vandersypen}}, \ and\
  \bibinfo {author} {\bibfnamefont {L.}~\bibnamefont {Kouwenhoven}},\ }in\
  \href@noop {} {\emph {\bibinfo {booktitle} {IEDM Technical Digest. IEEE
  International Electron Devices Meeting, 2004.}}}\ (\bibinfo {organization}
  {IEEE},\ \bibinfo {year} {2004})\ pp.\ \bibinfo {pages}
  {533--536}\BibitemShut {NoStop}%
\bibitem [{\citenamefont {Fujita}\ \emph {et~al.}(2019)\citenamefont {Fujita},
  \citenamefont {Morimoto}, \citenamefont {Kiyama}, \citenamefont {Allison},
  \citenamefont {Larsson}, \citenamefont {Ludwig}, \citenamefont {Valentin},
  \citenamefont {Wieck}, \citenamefont {Oiwa},\ and\ \citenamefont
  {Tarucha}}]{fujita2019angular}%
  \BibitemOpen
  \bibfield  {author} {\bibinfo {author} {\bibfnamefont {T.}~\bibnamefont
  {Fujita}}, \bibinfo {author} {\bibfnamefont {K.}~\bibnamefont {Morimoto}},
  \bibinfo {author} {\bibfnamefont {H.}~\bibnamefont {Kiyama}}, \bibinfo
  {author} {\bibfnamefont {G.}~\bibnamefont {Allison}}, \bibinfo {author}
  {\bibfnamefont {M.}~\bibnamefont {Larsson}}, \bibinfo {author} {\bibfnamefont
  {A.}~\bibnamefont {Ludwig}}, \bibinfo {author} {\bibfnamefont {S.~R.}\
  \bibnamefont {Valentin}}, \bibinfo {author} {\bibfnamefont {A.~D.}\
  \bibnamefont {Wieck}}, \bibinfo {author} {\bibfnamefont {A.}~\bibnamefont
  {Oiwa}}, \ and\ \bibinfo {author} {\bibfnamefont {S.}~\bibnamefont
  {Tarucha}},\ }\href@noop {} {\bibfield  {journal} {\bibinfo  {journal}
  {Nature communications}\ }\textbf {\bibinfo {volume} {10}},\ \bibinfo {pages}
  {1} (\bibinfo {year} {2019})}\BibitemShut {NoStop}%
\bibitem [{\citenamefont {Koppens}\ \emph {et~al.}(2006)\citenamefont
  {Koppens}, \citenamefont {Buizert}, \citenamefont {Tielrooij}, \citenamefont
  {Vink}, \citenamefont {Nowack}, \citenamefont {Meunier}, \citenamefont
  {Kouwenhoven},\ and\ \citenamefont {Vandersypen}}]{koppens2006driven}%
  \BibitemOpen
  \bibfield  {author} {\bibinfo {author} {\bibfnamefont {F.~H.}\ \bibnamefont
  {Koppens}}, \bibinfo {author} {\bibfnamefont {C.}~\bibnamefont {Buizert}},
  \bibinfo {author} {\bibfnamefont {K.-J.}\ \bibnamefont {Tielrooij}}, \bibinfo
  {author} {\bibfnamefont {I.~T.}\ \bibnamefont {Vink}}, \bibinfo {author}
  {\bibfnamefont {K.~C.}\ \bibnamefont {Nowack}}, \bibinfo {author}
  {\bibfnamefont {T.}~\bibnamefont {Meunier}}, \bibinfo {author} {\bibfnamefont
  {L.}~\bibnamefont {Kouwenhoven}}, \ and\ \bibinfo {author} {\bibfnamefont
  {L.}~\bibnamefont {Vandersypen}},\ }\href@noop {} {\bibfield  {journal}
  {\bibinfo  {journal} {Nature}\ }\textbf {\bibinfo {volume} {442}},\ \bibinfo
  {pages} {766} (\bibinfo {year} {2006})}\BibitemShut {NoStop}%
\bibitem [{\citenamefont {Press}\ \emph {et~al.}(2010)\citenamefont {Press},
  \citenamefont {De~Greve}, \citenamefont {McMahon}, \citenamefont {Ladd},
  \citenamefont {Friess}, \citenamefont {Schneider}, \citenamefont {Kamp},
  \citenamefont {H{\"o}fling}, \citenamefont {Forchel},\ and\ \citenamefont
  {Yamamoto}}]{press2010ultrafast}%
  \BibitemOpen
  \bibfield  {author} {\bibinfo {author} {\bibfnamefont {D.}~\bibnamefont
  {Press}}, \bibinfo {author} {\bibfnamefont {K.}~\bibnamefont {De~Greve}},
  \bibinfo {author} {\bibfnamefont {P.~L.}\ \bibnamefont {McMahon}}, \bibinfo
  {author} {\bibfnamefont {T.~D.}\ \bibnamefont {Ladd}}, \bibinfo {author}
  {\bibfnamefont {B.}~\bibnamefont {Friess}}, \bibinfo {author} {\bibfnamefont
  {C.}~\bibnamefont {Schneider}}, \bibinfo {author} {\bibfnamefont
  {M.}~\bibnamefont {Kamp}}, \bibinfo {author} {\bibfnamefont {S.}~\bibnamefont
  {H{\"o}fling}}, \bibinfo {author} {\bibfnamefont {A.}~\bibnamefont
  {Forchel}}, \ and\ \bibinfo {author} {\bibfnamefont {Y.}~\bibnamefont
  {Yamamoto}},\ }\href@noop {} {\bibfield  {journal} {\bibinfo  {journal}
  {Nature Photonics}\ }\textbf {\bibinfo {volume} {4}},\ \bibinfo {pages} {367}
  (\bibinfo {year} {2010})}\BibitemShut {NoStop}%
\bibitem [{\citenamefont {Press}\ \emph {et~al.}(2008)\citenamefont {Press},
  \citenamefont {Ladd}, \citenamefont {Zhang},\ and\ \citenamefont
  {Yamamoto}}]{press2008complete}%
  \BibitemOpen
  \bibfield  {author} {\bibinfo {author} {\bibfnamefont {D.}~\bibnamefont
  {Press}}, \bibinfo {author} {\bibfnamefont {T.~D.}\ \bibnamefont {Ladd}},
  \bibinfo {author} {\bibfnamefont {B.}~\bibnamefont {Zhang}}, \ and\ \bibinfo
  {author} {\bibfnamefont {Y.}~\bibnamefont {Yamamoto}},\ }\href@noop {}
  {\bibfield  {journal} {\bibinfo  {journal} {Nature}\ }\textbf {\bibinfo
  {volume} {456}},\ \bibinfo {pages} {218} (\bibinfo {year}
  {2008})}\BibitemShut {NoStop}%
\bibitem [{\citenamefont {Gupta}\ \emph {et~al.}(2001)\citenamefont {Gupta},
  \citenamefont {Knobel}, \citenamefont {Samarth},\ and\ \citenamefont
  {Awschalom}}]{gupta2001ultrafast}%
  \BibitemOpen
  \bibfield  {author} {\bibinfo {author} {\bibfnamefont {J.}~\bibnamefont
  {Gupta}}, \bibinfo {author} {\bibfnamefont {R.}~\bibnamefont {Knobel}},
  \bibinfo {author} {\bibfnamefont {N.}~\bibnamefont {Samarth}}, \ and\
  \bibinfo {author} {\bibfnamefont {D.}~\bibnamefont {Awschalom}},\ }\href@noop
  {} {\bibfield  {journal} {\bibinfo  {journal} {Science}\ }\textbf {\bibinfo
  {volume} {292}},\ \bibinfo {pages} {2458} (\bibinfo {year}
  {2001})}\BibitemShut {NoStop}%
\bibitem [{\citenamefont {Wei}\ and\ \citenamefont
  {Deng}(2014)}]{wei2014universal}%
  \BibitemOpen
  \bibfield  {author} {\bibinfo {author} {\bibfnamefont {H.-R.}\ \bibnamefont
  {Wei}}\ and\ \bibinfo {author} {\bibfnamefont {F.-G.}\ \bibnamefont {Deng}},\
  }\href@noop {} {\bibfield  {journal} {\bibinfo  {journal} {Optics express}\
  }\textbf {\bibinfo {volume} {22}},\ \bibinfo {pages} {593} (\bibinfo {year}
  {2014})}\BibitemShut {NoStop}%
\bibitem [{\citenamefont {Rosenblum}\ \emph {et~al.}(2018)\citenamefont
  {Rosenblum}, \citenamefont {Gao}, \citenamefont {Reinhold}, \citenamefont
  {Wang}, \citenamefont {Axline}, \citenamefont {Frunzio}, \citenamefont
  {Girvin}, \citenamefont {Jiang}, \citenamefont {Mirrahimi}, \citenamefont
  {Devoret} \emph {et~al.}}]{rosenblum2018cnot}%
  \BibitemOpen
  \bibfield  {author} {\bibinfo {author} {\bibfnamefont {S.}~\bibnamefont
  {Rosenblum}}, \bibinfo {author} {\bibfnamefont {Y.~Y.}\ \bibnamefont {Gao}},
  \bibinfo {author} {\bibfnamefont {P.}~\bibnamefont {Reinhold}}, \bibinfo
  {author} {\bibfnamefont {C.}~\bibnamefont {Wang}}, \bibinfo {author}
  {\bibfnamefont {C.~J.}\ \bibnamefont {Axline}}, \bibinfo {author}
  {\bibfnamefont {L.}~\bibnamefont {Frunzio}}, \bibinfo {author} {\bibfnamefont
  {S.~M.}\ \bibnamefont {Girvin}}, \bibinfo {author} {\bibfnamefont
  {L.}~\bibnamefont {Jiang}}, \bibinfo {author} {\bibfnamefont
  {M.}~\bibnamefont {Mirrahimi}}, \bibinfo {author} {\bibfnamefont {M.~H.}\
  \bibnamefont {Devoret}},  \emph {et~al.},\ }\href@noop {} {\bibfield
  {journal} {\bibinfo  {journal} {Nature communications}\ }\textbf {\bibinfo
  {volume} {9}},\ \bibinfo {pages} {652} (\bibinfo {year} {2018})}\BibitemShut
  {NoStop}%
\bibitem [{\citenamefont {Castelano}\ \emph {et~al.}(2018)\citenamefont
  {Castelano}, \citenamefont {de~Lima}, \citenamefont {Madureira},
  \citenamefont {Degani},\ and\ \citenamefont
  {Maialle}}]{castelano2018optimal}%
  \BibitemOpen
  \bibfield  {author} {\bibinfo {author} {\bibfnamefont {L.~K.}\ \bibnamefont
  {Castelano}}, \bibinfo {author} {\bibfnamefont {E.~F.}\ \bibnamefont
  {de~Lima}}, \bibinfo {author} {\bibfnamefont {J.~R.}\ \bibnamefont
  {Madureira}}, \bibinfo {author} {\bibfnamefont {M.~H.}\ \bibnamefont
  {Degani}}, \ and\ \bibinfo {author} {\bibfnamefont {M.~Z.}\ \bibnamefont
  {Maialle}},\ }\href@noop {} {\bibfield  {journal} {\bibinfo  {journal}
  {Physical Review B}\ }\textbf {\bibinfo {volume} {97}},\ \bibinfo {pages}
  {235301} (\bibinfo {year} {2018})}\BibitemShut {NoStop}%
\bibitem [{\citenamefont {Wang}\ \emph {et~al.}(2013)\citenamefont {Wang},
  \citenamefont {Wen}, \citenamefont {Zhu}, \citenamefont {Zhang},\ and\
  \citenamefont {Yeon}}]{wang2013deterministic}%
  \BibitemOpen
  \bibfield  {author} {\bibinfo {author} {\bibfnamefont {H.-F.}\ \bibnamefont
  {Wang}}, \bibinfo {author} {\bibfnamefont {J.-J.}\ \bibnamefont {Wen}},
  \bibinfo {author} {\bibfnamefont {A.-D.}\ \bibnamefont {Zhu}}, \bibinfo
  {author} {\bibfnamefont {S.}~\bibnamefont {Zhang}}, \ and\ \bibinfo {author}
  {\bibfnamefont {K.-H.}\ \bibnamefont {Yeon}},\ }\href@noop {} {\bibfield
  {journal} {\bibinfo  {journal} {Physics Letters A}\ }\textbf {\bibinfo
  {volume} {377}},\ \bibinfo {pages} {2870} (\bibinfo {year}
  {2013})}\BibitemShut {NoStop}%
\bibitem [{\citenamefont {Plantenberg}\ \emph {et~al.}(2007)\citenamefont
  {Plantenberg}, \citenamefont {De~Groot}, \citenamefont {Harmans},\ and\
  \citenamefont {Mooij}}]{plantenberg2007demonstration}%
  \BibitemOpen
  \bibfield  {author} {\bibinfo {author} {\bibfnamefont {J.}~\bibnamefont
  {Plantenberg}}, \bibinfo {author} {\bibfnamefont {P.}~\bibnamefont
  {De~Groot}}, \bibinfo {author} {\bibfnamefont {C.}~\bibnamefont {Harmans}}, \
  and\ \bibinfo {author} {\bibfnamefont {J.}~\bibnamefont {Mooij}},\
  }\href@noop {} {\bibfield  {journal} {\bibinfo  {journal} {Nature}\ }\textbf
  {\bibinfo {volume} {447}},\ \bibinfo {pages} {836} (\bibinfo {year}
  {2007})}\BibitemShut {NoStop}%
\bibitem [{\citenamefont {Imamog}\ \emph {et~al.}(1999)\citenamefont {Imamog},
  \citenamefont {Awschalom}, \citenamefont {Burkard}, \citenamefont
  {DiVincenzo}, \citenamefont {Loss}, \citenamefont {Sherwin}, \citenamefont
  {Small} \emph {et~al.}}]{imamog1999quantum}%
  \BibitemOpen
  \bibfield  {author} {\bibinfo {author} {\bibfnamefont {A.}~\bibnamefont
  {Imamog}}, \bibinfo {author} {\bibfnamefont {D.~D.}\ \bibnamefont
  {Awschalom}}, \bibinfo {author} {\bibfnamefont {G.}~\bibnamefont {Burkard}},
  \bibinfo {author} {\bibfnamefont {D.~P.}\ \bibnamefont {DiVincenzo}},
  \bibinfo {author} {\bibfnamefont {D.}~\bibnamefont {Loss}}, \bibinfo {author}
  {\bibfnamefont {M.}~\bibnamefont {Sherwin}}, \bibinfo {author} {\bibfnamefont
  {A.}~\bibnamefont {Small}},  \emph {et~al.},\ }\href@noop {} {\bibfield
  {journal} {\bibinfo  {journal} {Physical review letters}\ }\textbf {\bibinfo
  {volume} {83}},\ \bibinfo {pages} {4204} (\bibinfo {year}
  {1999})}\BibitemShut {NoStop}%
\bibitem [{\citenamefont {Monroe}\ \emph {et~al.}(1995)\citenamefont {Monroe},
  \citenamefont {Meekhof}, \citenamefont {King}, \citenamefont {Itano},\ and\
  \citenamefont {Wineland}}]{monroe1995demonstration}%
  \BibitemOpen
  \bibfield  {author} {\bibinfo {author} {\bibfnamefont {C.}~\bibnamefont
  {Monroe}}, \bibinfo {author} {\bibfnamefont {D.}~\bibnamefont {Meekhof}},
  \bibinfo {author} {\bibfnamefont {B.}~\bibnamefont {King}}, \bibinfo {author}
  {\bibfnamefont {W.~M.}\ \bibnamefont {Itano}}, \ and\ \bibinfo {author}
  {\bibfnamefont {D.~J.}\ \bibnamefont {Wineland}},\ }\href@noop {} {\bibfield
  {journal} {\bibinfo  {journal} {Physical review letters}\ }\textbf {\bibinfo
  {volume} {75}},\ \bibinfo {pages} {4714} (\bibinfo {year}
  {1995})}\BibitemShut {NoStop}%
\bibitem [{\citenamefont {O'Brien}\ \emph {et~al.}(2003)\citenamefont
  {O'Brien}, \citenamefont {Pryde}, \citenamefont {White}, \citenamefont
  {Ralph},\ and\ \citenamefont {Branning}}]{o2003demonstration}%
  \BibitemOpen
  \bibfield  {author} {\bibinfo {author} {\bibfnamefont {J.~L.}\ \bibnamefont
  {O'Brien}}, \bibinfo {author} {\bibfnamefont {G.~J.}\ \bibnamefont {Pryde}},
  \bibinfo {author} {\bibfnamefont {A.~G.}\ \bibnamefont {White}}, \bibinfo
  {author} {\bibfnamefont {T.~C.}\ \bibnamefont {Ralph}}, \ and\ \bibinfo
  {author} {\bibfnamefont {D.}~\bibnamefont {Branning}},\ }\href@noop {}
  {\bibfield  {journal} {\bibinfo  {journal} {Nature}\ }\textbf {\bibinfo
  {volume} {426}},\ \bibinfo {pages} {264} (\bibinfo {year}
  {2003})}\BibitemShut {NoStop}%
\bibitem [{\citenamefont {Burkard}\ \emph {et~al.}(1999)\citenamefont
  {Burkard}, \citenamefont {Loss},\ and\ \citenamefont
  {DiVincenzo}}]{burkard1999coupled}%
  \BibitemOpen
  \bibfield  {author} {\bibinfo {author} {\bibfnamefont {G.}~\bibnamefont
  {Burkard}}, \bibinfo {author} {\bibfnamefont {D.}~\bibnamefont {Loss}}, \
  and\ \bibinfo {author} {\bibfnamefont {D.~P.}\ \bibnamefont {DiVincenzo}},\
  }\href@noop {} {\bibfield  {journal} {\bibinfo  {journal} {Physical Review
  B}\ }\textbf {\bibinfo {volume} {59}},\ \bibinfo {pages} {2070} (\bibinfo
  {year} {1999})}\BibitemShut {NoStop}%
\bibitem [{\citenamefont {Elzerman}\ \emph {et~al.}(2004)\citenamefont
  {Elzerman}, \citenamefont {Hanson}, \citenamefont {Van~Beveren},
  \citenamefont {Witkamp}, \citenamefont {Vandersypen},\ and\ \citenamefont
  {Kouwenhoven}}]{elzerman2004single}%
  \BibitemOpen
  \bibfield  {author} {\bibinfo {author} {\bibfnamefont {J.}~\bibnamefont
  {Elzerman}}, \bibinfo {author} {\bibfnamefont {R.}~\bibnamefont {Hanson}},
  \bibinfo {author} {\bibfnamefont {L.~W.}\ \bibnamefont {Van~Beveren}},
  \bibinfo {author} {\bibfnamefont {B.}~\bibnamefont {Witkamp}}, \bibinfo
  {author} {\bibfnamefont {L.}~\bibnamefont {Vandersypen}}, \ and\ \bibinfo
  {author} {\bibfnamefont {L.~P.}\ \bibnamefont {Kouwenhoven}},\ }\href@noop {}
  {\bibfield  {journal} {\bibinfo  {journal} {nature}\ }\textbf {\bibinfo
  {volume} {430}},\ \bibinfo {pages} {431} (\bibinfo {year}
  {2004})}\BibitemShut {NoStop}%
\bibitem [{\citenamefont {Kloeffel}\ and\ \citenamefont
  {Loss}(2013)}]{kloeffel2013prospects}%
  \BibitemOpen
  \bibfield  {author} {\bibinfo {author} {\bibfnamefont {C.}~\bibnamefont
  {Kloeffel}}\ and\ \bibinfo {author} {\bibfnamefont {D.}~\bibnamefont
  {Loss}},\ }\href@noop {} {\bibfield  {journal} {\bibinfo  {journal} {Annu.
  Rev. Condens. Matter Phys.}\ }\textbf {\bibinfo {volume} {4}},\ \bibinfo
  {pages} {51} (\bibinfo {year} {2013})}\BibitemShut {NoStop}%
\bibitem [{\citenamefont {Steck}(2007)}]{steck2007quantum}%
  \BibitemOpen
  \bibfield  {author} {\bibinfo {author} {\bibfnamefont {D.~A.}\ \bibnamefont
  {Steck}},\ }\href@noop {} {\emph {\bibinfo {title} {Quantum and atom
  optics}}},\ Vol.~\bibinfo {volume} {47}\ (\bibinfo  {publisher} {Open
  Publication License, v1.0},\ \bibinfo {year} {2007})\ pp.\ \bibinfo {pages}
  {177--182}\BibitemShut {NoStop}%
\bibitem [{\citenamefont {Torrey}(1949)}]{torrey1949transient}%
  \BibitemOpen
  \bibfield  {author} {\bibinfo {author} {\bibfnamefont {H.}~\bibnamefont
  {Torrey}},\ }\href@noop {} {\bibfield  {journal} {\bibinfo  {journal}
  {Physical Review}\ }\textbf {\bibinfo {volume} {76}},\ \bibinfo {pages}
  {1059} (\bibinfo {year} {1949})}\BibitemShut {NoStop}%
\bibitem [{\citenamefont {Harper}\ and\ \citenamefont
  {Wherrett}(1977)}]{harper1977nonlinear}%
  \BibitemOpen
  \bibfield  {author} {\bibinfo {author} {\bibfnamefont {P.~G.}\ \bibnamefont
  {Harper}}\ and\ \bibinfo {author} {\bibfnamefont {B.~S.}\ \bibnamefont
  {Wherrett}},\ }\href@noop {} {\emph {\bibinfo {title} {Nonlinear Optics:
  Proceedings of the Sixteenth Scottish Universities Summer School in Physics,
  1975}}}\ (\bibinfo  {publisher} {Academic Pr},\ \bibinfo {year} {1977})\ pp.\
  \bibinfo {pages} {307--363}\BibitemShut {NoStop}%
\bibitem [{\citenamefont {Zajac}\ \emph {et~al.}(2018)\citenamefont {Zajac},
  \citenamefont {Sigillito}, \citenamefont {Russ}, \citenamefont {Borjans},
  \citenamefont {Taylor}, \citenamefont {Burkard},\ and\ \citenamefont
  {Petta}}]{zajac2018resonantly}%
  \BibitemOpen
  \bibfield  {author} {\bibinfo {author} {\bibfnamefont {D.~M.}\ \bibnamefont
  {Zajac}}, \bibinfo {author} {\bibfnamefont {A.~J.}\ \bibnamefont
  {Sigillito}}, \bibinfo {author} {\bibfnamefont {M.}~\bibnamefont {Russ}},
  \bibinfo {author} {\bibfnamefont {F.}~\bibnamefont {Borjans}}, \bibinfo
  {author} {\bibfnamefont {J.~M.}\ \bibnamefont {Taylor}}, \bibinfo {author}
  {\bibfnamefont {G.}~\bibnamefont {Burkard}}, \ and\ \bibinfo {author}
  {\bibfnamefont {J.~R.}\ \bibnamefont {Petta}},\ }\href@noop {} {\bibfield
  {journal} {\bibinfo  {journal} {Science}\ }\textbf {\bibinfo {volume}
  {359}},\ \bibinfo {pages} {439} (\bibinfo {year} {2018})}\BibitemShut
  {NoStop}%
\bibitem [{\citenamefont {Bengtsson}\ and\ \citenamefont
  {{\.Z}yczkowski}(2017)}]{bengtsson2017geometry}%
  \BibitemOpen
  \bibfield  {author} {\bibinfo {author} {\bibfnamefont {I.}~\bibnamefont
  {Bengtsson}}\ and\ \bibinfo {author} {\bibfnamefont {K.}~\bibnamefont
  {{\.Z}yczkowski}},\ }\href@noop {} {\emph {\bibinfo {title} {Geometry of
  quantum states: an introduction to quantum entanglement}}}\ (\bibinfo
  {publisher} {Cambridge university press},\ \bibinfo {year}
  {2017})\BibitemShut {NoStop}%
\end{thebibliography}%
\end{document}